\documentclass[lettersize,journal]{IEEEtran}
\usepackage{amsmath,amsfonts}
\usepackage{algorithmic}
\usepackage{algorithm}
\usepackage{array}
\usepackage[caption=false,font=scriptsize,labelfont=rm,textfont=rm]{subfig}
\usepackage{textcomp}
\usepackage{stfloats}
\usepackage{url}
\usepackage{verbatim}
\usepackage{graphicx}
\usepackage{cite}
\usepackage{bm}			
\usepackage{amssymb}		
\usepackage{mathrsfs}			
\usepackage{color}
\usepackage[colorlinks=true,citecolor=blue,linkcolor=blue,urlcolor=blue]{hyperref}

\allowdisplaybreaks
\begin{document}

\title{ Sum-Rate Maximization for Movable-Antenna Array Enhanced Downlink NOMA Systems }

\author{Nianzu Li,~Peiran Wu,~\IEEEmembership{Member,~IEEE},~Lipeng Zhu,~\IEEEmembership{Member,~IEEE},~Weidong Mei,~\IEEEmembership{Member,~IEEE},\\~Boyu Ning,~\IEEEmembership{Member,~IEEE},~and~Derrick Wing Kwan Ng,~\IEEEmembership{Fellow,~IEEE}\vspace{-0.2cm}

\thanks{Part of this work will be presented in 2025 IEEE International Conference on Communications Workshops \cite{ref49}.}
\thanks{N. Li and P. Wu are with the School of Electronics and Information Technology, Sun Yat-sen University, Guangzhou 510006, China (e-mail: linz5@mail2.sysu.edu.cn; wupr3@mail.sysu.edu.cn).}
\thanks{L. Zhu is with the Department of Electrical and Computer Engineering, National University of Singapore, Singapore 117583, Singapore (e-mail: zhulp@nus.edu.sg).}
\thanks{W. Mei and B. Ning are with the National Key Laboratory of Wireless Communications, University of Electronic Science and Technology of China, Chengdu 611731, China (e-mail: wmei@uestc.edu.cn; boydning@outlook.com).}
\thanks{D. W. K. Ng is with the School of Electrical Engineering and Telecommunications, University of New South Wales, Sydney, NSW 2052, Australia (e-mail: w.k.ng@unsw.edu.au).}
}

\markboth{Journal of \LaTeX\ Class Files,~Vol.~18, No.~9, September~2020}%
{Shell \MakeLowercase{\textit{et al.}}: A Sample Article Using IEEEtran.cls for IEEE Journals}


\maketitle

\begin{abstract}
Movable antenna (MA) systems have recently attracted significant attention in the field of wireless communications owing to their exceptional capability to proactively reconfigure wireless channels via flexible antenna movements. In this paper, we investigate the resource allocation design for an MA array-enhanced downlink non-orthogonal multiple access (NOMA) system, where a base station deploys multiple MAs to serve multiple single-antenna users. Our goal is to maximize the sum rate of all users by jointly optimizing the transmit beamforming, positions of MAs, successive interference cancellation (SIC) decoding order, and users' corresponding decoding indicator matrix, while adhering to constraints on the maximum transmit power and finite MA moving region. The formulated problem is inherently highly non-convex, rendering it challenging to acquire a globally optimal solution. As a compromise, we propose a low-complexity two-stage optimization algorithm to obtain an effective suboptimal solution. Specifically, in stage one, the SIC decoding order is first determined by solving a channel gain maximization problem. Then, in stage two, with the given SIC decoding order, the beamforming vectors, MA positions, and users' decoding indicator matrix are iteratively optimized by capitalizing on alternating optimization, successive convex approximation (SCA), and genetic algorithm (GA). Simulation results unveil that the sum-rate performance of the proposed MA-enabled downlink NOMA system significantly outperforms that of conventional fixed-position antenna (FPA) systems. Moreover, the results also show that the antenna position optimization in the proposed algorithm can further enhance the advantages of NOMA over space division multiple access (SDMA).
\end{abstract}


\section{Introduction}
\IEEEPARstart{W}{ith} rapid advancements in wireless communication applications, such as the Internet-of-Things (IoT) devices, ultra-high-definition video streaming, mobile edge computing etc., global anticipation for the forthcoming sixth-generation (6G) wireless network has dramatically burgeoned\cite{ref51}. Specifically, these networks are poised to support higher data rates, ultra-low latency, massive access, and simultaneous connectivity for a vast number of devices\cite{ref52}. 
Motivated by these evolving demands, numerous advanced communication technologies have emerged over the past few decades. For instance, multiple-input multiple-output (MIMO) technology has evolved from single-input single-output (SISO) systems, enabling multi-stream parallel data transmission between transceivers and effectively enhancing systems' spectral efficiency\cite{ref51,ref52}. Besides, as wireless communication systems migrate toward operating in higher frequency bands, e.g., millimeter-wave (mmWave) and terahertz (THz) bands, MIMO systems with large-scale antenna arrays, commonly referred to as massive MIMO systems, have been proposed. Compared with conventional MIMO, massive MIMO leverages substantially increased array gains to effectively combat severe path loss, while exploiting its rich spatial degrees of freedom to facilitate spatial multiplexing\cite{ref53}. Apart from these developments, another promising technology in wireless communications is non-orthogonal multiple access (NOMA), which enables different users to simultaneously share the same time-frequency-code resources for improving spectral efficiency and system throughput\cite{ref48}. In particular, by resorting to the successive interference cancellation (SIC) technique, NOMA enables each user to sequentially remove co-channel interference caused by other users with prior decoding orders before decoding its own information. Consequently, NOMA provides an effective approach to distinguish multi-user signals and satisfy the stringent requirements for both massive connectivity and service quality, especially in practical scenarios with scarce spectrum availability.

Despite these significant technological advancements, the aforementioned technologies are all established based upon traditional fixed-position antenna (FPA) systems, which may encounter severe limitations in fulfilling the stringent requirements of future wireless networks. Specifically, due to their static and discrete antenna deployments, FPA systems inherently experience random and uncontrollable wireless channel fading, rendering them incapable of fully exploiting continuous channel variations across spatial regions\cite{ref54}. Therefore, channel conditions at these fixed locations may severely degrade under unfavorable scenarios, such as ``deep fading'', which dramatically deteriorates overall system performance.
Fortunately, the above limitations are expected to be addressed with the emergence of the recently developed movable antenna (MA) technology\cite{ref1},\cite{ref35}. In fact, the earliest work investigating MA-enabled wireless communications can be traced to \cite{ref72} published in 2009, where the spatial diversity gain of a single MA was characterized. Then, this technology has attracted increasing attention in the communication community since 2022, showing prosperous development trends together with other reconfigurable antenna technologies, such as fluid antenna system (FAS)\cite{ref75},\cite{ref71}, and pinching antenna (PA)\cite{ref55},\cite{ref69}. It is  worth noting that although the manufacturing techniques for MA and FAS may differ in practice, e.g., MA typically employs mechanically- or electronically-driven components, while FAS is fabricated by utilizing conductive fluids or reconfigurable pixels, they share a similar capability of flexible antenna repositioning. In particular, by strategically adjusting antenna positions within a localized region at the transceivers, they can effectively exploit the spatial variations of wireless channels to establish more favorable propagation conditions. Moreover, the antenna position optimization in an MA array can efficiently reconfigure array geometry and thus yields more flexible beamforming\cite{ref28}. In addition, the PA technology focuses on combating the large-scale fading of wireless channels. Specifically, by employing a long dielectric waveguide as transmission medium, small dielectric particles can be proactively moved or selectively activated to radiate electromagnetic signals at different positions along the one-dimensional waveguide. Thus, it achieves an equivalent effect of large-scale antenna movement, facilitating the establishment of a strong line-of-sight propagation link between transceivers and effectively reducing the impacts of large-scale path loss\cite{ref79}. In summary, these reconfigurable antenna technologies provide a groundbreaking approach to proactively reconfigure wireless channels by dynamically altering antenna positions, introducing additional degrees of freedom (DoFs) to enhance overall performance of communication systems.

The substantial benefits of MA/FAS/PA have ignited extensive research interest in integrating them into various wireless communication applications. For instance, the authors of \cite{ref3} presented a theoretical analysis of the signal-to-noise ratio (SNR) gain in an MA-enabled SISO communication system. Their results demonstrate that by adaptively repositioning the transmit antenna to cope with the dynamically-varying nature of wireless channel conditions, the system performance can be significantly improved compared with its FPA counterpart. Besides, the authors of \cite{ref57} analyzed the outage probability and diversity gain in FAS-aided SISO communication systems and proposed a suboptimal FAS architecture with $N$ ports. Their results reveal that this simple solution, despite relying on only one radio frequency (RF) chain, can achieve performance comparable to multiple RF chain-based systems. Moreover, in \cite{ref27,ref73}, the authors investigated an MA-assisted multiple-input single-output (MISO) single-user communication system and proposed a graph-based approach to acquire optimal discrete MA positions for maximizing the received signal's power. In addition, the authors of \cite{ref56} studied a PA-assisted MISO communication system and developed a two-stage algorithm for maximizing the downlink transmission rate. Furthermore, the deployment of MAs in MIMO communication systems was investigated in \cite{ref4}, wherein the authors maximized the channel capacity by jointly designing the transmit/receive MA positions and the transmit covariance matrix. 


Meanwhile, the applications of MA/FAS/PA have also been widely explored in more general multi-user communication scenarios. For example, the authors in \cite{ref9}, \cite{ref58}, \cite{ref61} explored MA-enabled multi-user uplink transmission systems, aiming to minimize the total transmit power of the base station (BS), maximize the network capacity, and maximize the users' ergodic sum rate, respectively. On the other hand, there have also been several research efforts for the downlink transmission scenarios. Specifically, in \cite{ref74,ref59}, the authors investigated an MA/FAS-assisted integrated sensing and communications (ISAC) downlink multi-user MIMO system; while in \cite{ref60}, the authors explored downlink beamforming in a PA-assisted multi-user MIMO system. The above works on MA/FAS/PA-aided multi-user communications have focused on traditional orthogonal multiple access (OMA) or space division multiple access (SDMA) systems. Furthermore, MAs/FASs/PAs have also been integrated into other wireless system setups, such as over-the-air computation \cite{ref36}, multi-cast transmissions \cite{ref39}, and physical layer security \cite{ref78},\cite{ref76},\cite{ref63},\cite{ref77}.


In this paper, we aim to advance the integration of MA in NOMA systems. As previously discussed, since both NOMA and MA possess an exceptional capability to enhance wireless network's performance, i.e., improving spectral efficiency by making full use of spectral resources and channel variations, respectively, there is a significant potential to synergistically integrate their advantages to further improve network communication. To this end, several existing works have demonstrated the significant performance benefits achieved by  combining MA with NOMA\cite{ref41}, \cite{ref42}, \cite{ref2}. However, in these works, only a single MA is considered to be deployed at the user side, whereas in practice, there are typically more available energy and space at the BS side for flexible antenna movements. Besides, it is worth emphasizing that the problem of exploiting MAs at the BS side is fundamentally different from that at the user side. Specifically, for antenna positioning design at the BS, each MA needs to accommodate the trade-offs among the channel conditions of all users. In contrast, for antenna positioning design at the user side, each user can reconfigure its individual channel condition by  strategically repositioning its own MA. As such, the results in \cite{ref41}, \cite{ref42}, \cite{ref2} cannot be directly applied to scenarios involving the deployment of MAs at the BS for enhancing communication performance of NOMA systems. Furthermore, for downlink NOMA systems, the works in \cite{ref41} and \cite{ref42} did not address the SIC decoding ordering problem, which becomes particularly critical in MA-enabled  NOMA systems due to dynamically-varying channel conditions resulting from antenna movements.

To fill in the above gaps, in this paper, we explore a more general and practical scenario for an MA-enhanced downlink NOMA system, where multiple MAs are deployed at the BS to serve multiple single-FPA users. In particular, we investigate the resource allocation problem of BS transmit beamforming, MA positioning, SIC ordering, and adaptive user decoding for maximizing the system's sum rate. The main contributions of this paper are summarized as follows:
\begin{itemize}
	\item[1)] We propose to deploy an MA array at the BS to enhance the performance of a downlink multi-user MISO-NOMA system. First, the field-response based channel model is employed to characterize the multi-path channel response between the BS and each single-antenna user as a function of the antenna position vector (APV). Subsequently, we formulate an optimization problem to maximize the system's sum rate by jointly optimizing the beamforming vectors of the BS, the APV of all MAs, the SIC decoding order, and the users' decoding indicator matrix, while satisfying the transmit power budget of the BS and finite MA moving region constraints.
	
	\item[2)] As the formulated problem is inherently non-convex, we propose a two-stage iterative algorithm to acquire a suboptimal solution, which consists of a SIC decoding order determination stage followed by a joint optimization stage for transmit beamforming, MA positioning, and adaptive user decoding. Specifically, in the first stage, we determine the SIC decoding order by addressing an overall channel gain maximization problem. Next, in the second stage, given the SIC decoding order, we decouple the original optimization problem into several subproblems by invoking an alternating optimization framework and solve them iteratively until convergence. In particular, the optimization subproblems of the transmit beamforming vectors and each MA position are tackled by applying the successive convex approximation (SCA) technique, while that of the users' decoding indicator matrix is addressed by leveraging a genetic algorithm (GA).
	
	\item[3)] Finally, numerical simulations are conducted to evaluate the performance of the proposed MA-enhanced downlink NOMA system. It is shown that the proposed scheme can achieve substantial improvements over traditional FPA- and SDMA-based systems in terms of sum-rate performance. Besides, it is revealed that exploiting MAs in NOMA systems can further amplify the performance advantages of NOMA over conventional SDMA, verifying the great benefits of combining NOMA with MA to enhance communication networks. Furthermore, results also indicate that the SIC decoding order obtained by our proposed algorithm can achieve performance close to its exhaustive search counterpart, demonstrating the effectiveness of the derived suboptimal solution.
	
	
\end{itemize}

The rest of this paper is organized as follows. Section \ref{section_2} introduces the system model of the considered MA-enhanced downlink NOMA system and formulates the sum-rate maximization problem. Section \ref{section_3} elaborates on the proposed two-stage optimization algorithm for handling the formulated non-convex problem. Simulation results are provided in Section \ref{section_4} to evaluate the effectiveness of the proposed solution and this paper is finally concluded in Section \ref{section_5}.

\textit{Notations}: Boldface lowercase and uppercase letters denotes vectors and matrices, respectively. The operators $(\cdot)^\mathrm{T}$, $(\cdot)^{*}$, and $(\cdot)^\mathrm{H}$ denote transpose, conjugate, and conjugate transpose, respectively. $\mathbb{C}$ denotes the complex space and $\mathbb{C}^{M\times N}$ denotes the space of $M\times N$ matrices with complex-valued elements. $[\cdot]_i$ and $[\cdot]_{i,j}$ denote the $i$-th element of a vector and the $(i,j)$-th element of a matrix, respectively. $|x|$ and $\mathrm{arg}(x)$ denote the absolute value and phase of the complex number $x$, respectively. $\|\mathbf{a}\|_2$ denotes the Euclidean norm of the vector $\mathbf{a}$. $\|\mathbf{A}\|_2$, $\|\mathbf{A}\|_{\mathrm{F}}$, $\mathrm{rank}(\mathbf{A})$, and $\mathrm{Tr}(\mathbf{A})$ denote the spectral norm, Frobenius norm, rank, and trace of the matrix $\mathbf{A}$, respectively. $\mathbf{A}\succeq\mathbf{0}$ indicates that the matrix $\mathbf{A}$ is positive semi-definite. $\mathbf{I}_N$ denotes the identity matrix with dimension $N$. The gradient and Hessian matrix of a function $f(\cdot)$ are denoted by $\nabla f(\cdot)$ and $\nabla^2 f(\cdot)$, respectively. $\mathcal{CN}(0,\sigma^2)$ denotes the circularly symmetric complex Gaussian distribution with zero mean and variance $\sigma^2$. $\mathbb{E}\{\cdot\}$ denotes the expectation operator.

\section{System Model and Problem Formulation}\label{section_2}
As shown in Fig. 1, we consider an MA-enhanced downlink MISO system employing NOMA, where a BS is equipped with $M$ transmit MAs to serve $K$ single-antenna users. The set of all users is denoted by $\mathcal{K}=\left\{1,2,\dots,K\right\}$ and the set of all MAs is denoted by $\mathcal{M}=\left\{1,2,\dots,M\right\}$. Each MA at the BS is individually connected to
an RF chain via a flexible cable, enabling dynamic repositioning in a localized two-dimensional (2D) region, $\mathcal{C}_t$, to proactively improve channel conditions\cite{ref1,ref3}. Specifically, the position of the $m$-th MA is represented by the 2D Cartesian coordinates, i.e., $\mathbf{u}_m=\left[x_m,y_m\right]^{\mathrm{T}}\in\mathcal{C}_t,\forall m\in\mathcal{M}$. Without loss of generality, we assume that $\mathcal{C}_t$ is a square region with side length $A$.

Let $\mathbf{h}_k(\tilde{\mathbf{u}})\in\mathbb{C}^{M\times1}$ denote the channel vector between the $M$ MAs at the BS and user $k$, with $\tilde{\mathbf{u}}=\left[\mathbf{u}_1^{\mathrm{T}},\mathbf{u}_2^{\mathrm{T}},\dots,\mathbf{u}_M^{\mathrm{T}}\right]^{\mathrm{T}}$ denoting the APV for all MAs. Therefore, the received signal at user $k$ can be expressed as
\begin{align}
	\label{eq1}
	y_k=\mathbf{h}_k^{\mathrm{H}}(\tilde{\mathbf{u}})\sum_{j=1}^{K}\mathbf{w}_js_j+n_k,
\end{align}
where $s_j\in\mathbb{C}$ denotes the information-bearing symbol for user $j$ with zero mean, unit variance, and $\mathbb{E}\left\{s_j^*s_q\right\}=0,\forall j\neq q$. Also, $\mathbf{w}_j\in\mathbb{C}^{M\times 1}$ represents the corresponding beamforming vector and $n_k\sim\mathcal{CN}\left(0,\sigma_k^2\right)$ is the zero-mean additive white Gaussian
noise (AWGN) at user $k$ with power $\sigma_k^2$. 
\begin{figure}[t]
	\centering
	\includegraphics[width=0.469\textwidth]{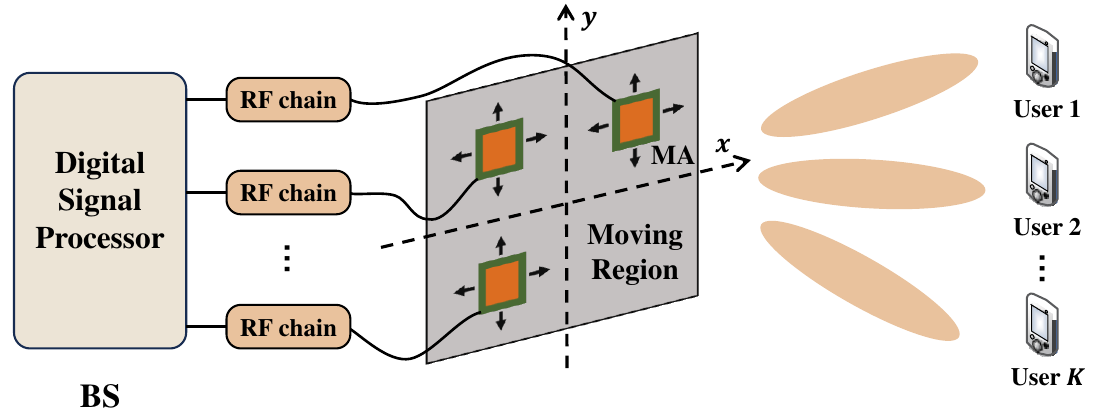}
	\caption{Illustration of the MA-enhanced downlink MISO-NOMA system.}
	\label{system_model}
\end{figure}

\subsection{Field-Response Based Channel Model}
In this paper, we employ the field-response based channel model under the far-field condition\cite{ref3}, where the channel response from the BS to each user is represented as the superposition of coefficients from multiple channel paths. Let $L_k,\forall k \in \mathcal{K}$, denote the number of transmit paths from the BS to user $k$. The elevation and azimuth angles of departure (AoDs) for the $\ell$-th transmit path between the BS and user $k$ are denoted as $\theta_{k,\ell}$ and $\phi_{k,\ell}$, respectively. Thus, the transmit field-response vector (FRV) for the channel between the $m$-th MA at the BS and user $k$ can be given by
\begin{align}
	\label{eq2}
	\mathbf{g}_k(\mathbf{u}_m)=\left[e^{\mathrm{j}\frac{2\pi}{\lambda}\rho_{k,1}(\mathbf{u}_m)},\dots,e^{\mathrm{j}\frac{2\pi}{\lambda}\rho_{k,L_k}(\mathbf{u}_m)}\right]^{\mathrm{T}},
\end{align}
where $\lambda$ denotes the signal carrier wavelength and $\rho_{k,\ell}(\mathbf{u}_m)=x_m\sin\theta_{k,\ell}\cos\phi_{k,\ell}+y_m\cos\theta_{k,\ell},1\leq\ell\leq L_k$, denotes the signal propagation path difference for the $\ell$-th channel path of user $k$ between the $m$-th MA's position $\mathbf{u}_m$ and the reference point at the BS, i.e., $\mathbf{o}=[0,0]^{\mathrm{T}}$. As a result, the channel vector between the BS and user $k$ can be expressed as
\begin{align}
	\label{eq3}
	\mathbf{h}_k(\tilde{\mathbf{u}})=\mathbf{G}_k^{\mathrm{H}}(\tilde{\mathbf{u}})\mathbf{f}_k,
\end{align}
where $\mathbf{G}_k(\tilde{\mathbf{u}})=\left[\mathbf{g}_k(\mathbf{u}_1),\dots,\mathbf{g}_k(\mathbf{u}_M)\right]\in\mathbb{C}^{L_k\times M}$ denotes the field-response matrix (FRM) for all MAs at the BS and $\mathbf{f}_k=\left[f_{k,1},f_{k,2},\dots,f_{k,L_k}\right]^{\mathrm{T}}\in\mathbb{C}^{L_k\times 1}$ denotes the path-response vector (PRV), representing the $L_k$ multi-path coefficients from the reference point at the BS to user $k$.

\subsection{Adaptive SIC Decoding Scheme}
According to the NOMA protocol, each user applies SIC to recover its intended signal. Let $z_k$ denote the user occupying the $k$-th position in the decoding sequence. Specifically, if $z_k=i$, it indicates that the signal intended for user $i$ will be the $k$-th one to be decoded. For conventional fixed SIC decoding schemes, each user is required to sequentially decode signals for other users with prior decoding orders\cite{ref41},\cite{ref38}. However, this fixed decoding architecture may generally incur substantial performance degradation since it does not consider the characteristics of potential inter-user channel correlation. A representative example arises when users exhibit mutually orthogonal channel vectors. In such a scenario, the optimal transmission strategy is to apply maximum ratio transmission (MRT) beamforming at the BS, which enables direct signal decoding at each user without the need for performing interference cancellation. In contrast, enforcing one user to decode the signal for other users may potentially deteriorate overall system performance\cite{ref70}.

To address the above problem, in this paper, we employ an adaptive SIC decoding scheme for each user\cite{ref42}. Specifically, we introduce a $K\times K$ decoding indicator matrix $\mathbf{\Pi}$, where each element is a binary optimization variable, denoted by $\pi_{z_i,z_j},1\leq i,j \leq K$, indicating whether the $j$-th user in the decoding sequence decodes the signal for the $i$-th user in the decoding sequence. For instance, if $\pi_{z_i,z_j}=1$, the $j$-th decoded user is required to decode the signal for the $i$-th decoded user; while if $\pi_{z_i,z_j}=0$, the $j$-th decoded user does not need to decode the signal intended for the $i$-th decoded user but treats it as interference. Based on this definition, the decoding indicator matrix $\mathbf{\Pi}$ can be expressed as
\begin{align}
	\label{eq4}
	\mathbf{\Pi}=\begin{bmatrix}
		1 & \pi_{z_1,z_2} & \pi_{z_1,z_3} & \cdots & \pi_{z_1,z_K}\\
		0 & 1 & \pi_{z_2,z_3} & \cdots & \pi_{z_2,z_K}\\
		\vdots & \vdots & \vdots & \ddots & \vdots\\
		0 & 0 & 0 & \cdots & 1
	\end{bmatrix},
\end{align}
which is an upper triangular matrix since the decoding order constrains that each user only decodes signals for itself and other users with prior decoding orders.
Thereby, the signal-to-interference-plus-noise ratio (SINR) of user $z_k$ to decode its own signal is given by
\begin{align}
	\label{eq5}
	&\gamma_{z_k\rightarrow z_k}=\notag\\
	&\frac{\left|\mathbf{h}_{z_k}^{\mathrm{H}}\left(\tilde{\mathbf{u}}\right)\mathbf{w}_{z_k}\right|^2}{\sum\limits_{j=1}^{K}\left|\mathbf{h}_{z_k}^{\mathrm{H}}\left(\tilde{\mathbf{u}}\right)\mathbf{w}_{z_j}\right|^2-\sum\limits_{j=1}^{k}\pi_{z_j,z_k}\left|\mathbf{h}_{z_k}^{\mathrm{H}}\left(\tilde{\mathbf{u}}\right)\mathbf{w}_{z_j}\right|^2+\sigma_{z_k}^2},
\end{align}
where the second term in the denominator is the inter-user interference removed by SIC. 
Besides, for any $z_k<z_i$ and $\pi_{z_k,z_i}=1$, the SINR of user $z_i$ to decode the signal for user $z_k$ is given by
\begin{align}
	\label{eq6}
	&\gamma_{z_k\rightarrow z_i}=\notag\\
	&\frac{\left|\mathbf{h}_{z_i}^{\mathrm{H}}\left(\tilde{\mathbf{u}}\right)\mathbf{w}_{z_k}\right|^2}{\sum\limits_{j=1}^{K}\left|\mathbf{h}_{z_i}^{\mathrm{H}}\left(\tilde{\mathbf{u}}\right)\mathbf{w}_{z_j}\right|^2-\sum\limits_{j=1}^{k}\pi_{z_j,z_i}\left|\mathbf{h}_{z_i}^{\mathrm{H}}\left(\tilde{\mathbf{u}}\right)\mathbf{w}_{z_j}\right|^2+\sigma_{z_i}^2}.
\end{align}
As a result, the achievable rate for decoding user $z_k$'s signal can be expressed as\cite{ref42}
\begin{align}
	\label{eq7}
	R_{k}=\min\nolimits_{\{\forall z_i | \pi_{z_k,z_i}=1\}}\log_2\big(1+\gamma_{z_k\rightarrow z_i}\big).
\end{align}

\subsection{Optimization Problem Formulation}
In this paper, we aim to maximize the sum rate of all users by jointly optimizing the APV of all MAs, the beamforming vectors at the BS, the SIC decoding order, and the decoding indicator matrix. To investigate the performance limits of the proposed MA-enhanced NOMA system, we assume that the channel state information (CSI) of all involved links is available at the BS for resource allocation design. Thus, the optimization problem can be formulated as
\begin{subequations}
	\label{eq8}
	\begin{align}
		\max_{\tilde{\mathbf{u}},\{\mathbf{w}_k,z_k\},\mathbf{\Pi}}& \quad \sum_{k=1}^{K}R_{k}\label{eq8a}\\
		\mathrm{s.t.}
		& \quad \mathbf{u}_m\in\mathcal{C}_t,~\forall m \in \mathcal{M},\label{eq8b}\\
		& \quad \|\mathbf{u}_m-\mathbf{u}_n\|_2\geq D,~\forall m,n\in\mathcal{M},m\neq n,\label{eq8c}\\
		& \quad \sum_{k=1}^{K}\|\mathbf{w}_k\|^2\leq P_{\max},\label{eq8d}\\
		& \quad R_{k}\geq R_k^{\min},~\forall k \in \mathcal{K},\label{eq8e}\\
		& \quad z_k\in\Omega,~\forall k\in\mathcal{K},\label{eq8f}\\
		& \quad \pi_{z_k,z_i}\in\left\{0,1\right\}, \forall k,i\in\mathcal{K},k< i,\label{eq8g}
	\end{align}
\end{subequations}
where constraint \eqref{eq8b} confines that each MA can only be moved within the given localized region, $\mathcal{C}_t$; constraint \eqref{eq8c} ensures the minimum inter-MA distance $D$ to avoid potential coupling effects among antennas; constraint \eqref{eq8d} ensures that the transmit power of the BS does not exceed its maximum value, $P_{\max}$; constraint \eqref{eq8e} ensures a minimum rate requirement $R_k^{\min}$ for each user to guarantee its quality-of-service (QoS); constraint \eqref{eq8f} represents the set of all possible SIC decoding orders, $\Omega$; and constraint \eqref{eq8g} indicates that the elements of the decoding indicator matrix are binary variables.

Problem \eqref{eq8} is challenging to solve optimally due to the following reasons. First, the objective in \eqref{eq8a} is highly non-concave with respect to the variables $\tilde{\mathbf{u}},\{\mathbf{w}_k\}$, $\{z_k\}$, and $\mathbf{\Pi}$. Besides, $\{z_k\}$ and $\mathbf{\Pi}$ are combinatorial optimization variables. Finally, the four sets of high-dimensional optimization variables are tightly coupled, rendering the problem more intractable. Thereby, existing optimization methods cannot be directly applied. In the following section, we will propose a computationally efficient algorithm to acquire a high-quality suboptimal solution.

\section{Proposed Solution}\label{section_3}
In this section, we develop a two-stage algorithm for efficiently solving problem \eqref{eq8}. The main idea is to decompose the original complicated optimization into two stages. First, the SIC decoding order is determined by addressing an overall channel gain maximization problem. Next, given the SIC decoding order, an alternating optimization framework is established to iteratively optimize the transmit beamforming vectors, MA positions, and decoding indicator matrix.

\subsection{SIC Decoding Order Determination}
Since NOMA is employed in this paper, the SIC decoding order constitutes an important design factor. In fact, an inappropriate decoding order may induce potential performance degradation. Note that for conventional FPA-based downlink NOMA systems, the SIC decoding order is typically obtained as the increasing order of users' channel gains\cite{ref48}. However, in the proposed MA-based downlink NOMA system, each user's channel gain is dynamically-varying due to antenna repositioning. Although an exhaustive search can be adopted to find the optimal decoding order, i.e., enumerating all possible solutions of $\{z_k\}$, its computational complexity is prohibitively high. To tackle this problem, we aim to maximize the overall channel gains between the BS and all users by optimizing the APV of the MAs, $\tilde{\mathbf{u}}$, and then determine the decoding order based on each user's individual channel gain\footnote{This intuitive approach can be justified by the fact that MAs typically play an important role in multi-user scenarios by improving each user's channel gain\cite{ref11}. As such, it is favorable to determine the SIC decoding order after maximizing the overall channel gains of all users. This strategy will be validated through simulations in Section \ref{section_4}.
 }. Thus, the optimization problem with respect to $\tilde{\mathbf{u}}$ can be formulated as
\begin{subequations}
	\label{eq9}
	\begin{align}
		\max_{\tilde{\mathbf{u}}}& \quad \sum_{k=1}^{K}\left\|\mathbf{h}_k\left(\tilde{\mathbf{u}}\right)\right\|_2^2 \label{eq9a}\\
		\mathrm{s.t.}
		& \quad \text{\eqref{eq8b}, \eqref{eq8c}}.
	\end{align}
\end{subequations}

Problem \eqref{eq9} is non-trivial to solve due to its non-concave objective function and the highly non-convex constraint \eqref{eq8c}. To resolve this issue, we employ an alternating optimization framework to iteratively optimize each MA position, i.e., $\mathbf{u}_m$, with the others, i.e., $\{\mathbf{u}_n,n\neq m\}$, being fixed by exploiting the SCA technique. First, it is worth noting that $\|\mathbf{h}_k(\tilde{\mathbf{u}})\|_2^2$ can be expanded as $\|\mathbf{h}_k(\tilde{\mathbf{u}})\|_2^2=\sum_{m=1}^{M}|\mathbf{g}_k^{\mathrm{H}}(\mathbf{u}_m)\mathbf{f}_k|^2$, where each term depends solely on a specific MA position. Therefore, for each MA position $\mathbf{u}_m$, the optimization problem can be rewritten as
\begin{subequations}
	\label{eq10}
	\begin{align}
		\max_{\mathbf{u}_m}& \quad \Phi(\mathbf{u}_m)=\sum_{k=1}^{K} \left|\mathbf{g}_k^{\mathrm{H}}(\mathbf{u}_m)\mathbf{f}_k\right|^2 \label{eq10a}\\
		\mathrm{s.t.}
		& \quad \mathbf{u}_m\in\mathcal{C}_t,\label{eq10b}\\
		& \quad \|\mathbf{u}_m-\mathbf{u}_n\|_2\geq D,~\forall n\neq m.\label{eq10c}
	\end{align}
\end{subequations}
Next, denote the gradient and Hessian matrix of $\Phi(\mathbf{u}_m)$ with respect to $\mathbf{u}_m$ as $\nabla\Phi(\mathbf{u}_m)$ and $\nabla^2\Phi(\mathbf{u}_m)$, respectively, with their closed-form expressions provided in Appendix \ref{appendix_A}.  
By constructing a positive real number $\delta_m$ such that $\delta_m\mathbf{I}_2\succeq\nabla^2\Phi(\mathbf{u}_m)$ and drawing on Taylor’s theorem, we can construct a concave surrogate function to globally lower-bound $\Phi(\mathbf{u}_m)$ with given local point $\mathbf{u}_m^t$ as\cite{ref4}
\begin{align}
	\label{eq11}
	\Phi\left(\mathbf{u}_m\right)\geq&\Phi\Big(\mathbf{u}_m^t\Big)+\nabla\Phi\Big(\mathbf{u}_m^t\Big)^{\mathrm{T}}\Big(\mathbf{u}_m-\mathbf{u}_m^t\Big)\notag\\
	&-\frac{\delta_m}{2}\Big(\mathbf{u}_m-\mathbf{u}_m^t\Big)^{\mathrm{T}}\Big(\mathbf{u}_m-\mathbf{u}_m^t\Big)\notag\\
	\triangleq&\Phi^{\mathrm{lb},t}(\mathbf{u}_m).
\end{align}
Note that since $\left\|\nabla^2\Phi(\mathbf{u}_m)\right\|_{\mathrm{F}}\mathbf{I}_2\succeq\left\|\nabla^2\Phi(\mathbf{u}_m)\right\|_2\mathbf{I}_2\succeq\nabla^2\Phi(\mathbf{u}_m)$, the real-valued  $\delta_{m}$ in \eqref{eq11} can be determined as
 \begin{align}
 	\label{eq12}
 	&\delta_{m}=\left\|\nabla^2\Phi(\mathbf{u}_m)\right\|_{\mathrm{F}}=\left[\left(\frac{\partial^2\Phi(\mathbf{u}_m)}{\partial x_m^2}\right)^2+\left(\frac{\partial^2\Phi(\mathbf{u}_m)}{\partial x_m\partial y_m}\right)^2\right.\notag\\
 	&\left.+\left(\frac{\partial^2\Phi(\mathbf{u}_m)}{\partial y_m\partial x_m}\right)^2+\left(\frac{\partial^2\Phi(\mathbf{u}_m)}{\partial y_m^2}\right)^2\right]^{\frac{1}{2}}.
 \end{align}
 
 Hereto, the remaining issue for tackling problem \eqref{eq10} lies in handling non-convex constraint \eqref{eq10c}. Note that this constraint is equivalent to $\|\mathbf{u}_m-\mathbf{u}_n\|_2^2\geq D^2, \forall m\neq n$. Since the term $\|\mathbf{u}_m-\mathbf{u}_n\|_2^2$ is convex with respect to $\mathbf{u}_m$, by applying the first-order Taylor series expansion, it is lower-bounded by
 \begin{align}
 	\label{eq13}
 	\big\|\mathbf{u}_m-\mathbf{u}_n\big\|_2^2\geq&\big\|\mathbf{u}_m^t-\mathbf{u}_n\big\|_2^2+2\Big(\mathbf{u}_m^t-\mathbf{u}_n\Big)^{\mathrm{T}}\Big(\mathbf{u}_m-\mathbf{u}_m^t\Big)\notag\\
 	\triangleq&U^{\mathrm{lb},t}_n(\mathbf{u}_m),
 \end{align}
based on which a convex subset of constraint \eqref{eq10c} can be established as
\begin{align}
	\label{eq14}
	U^{\mathrm{lb},t}_n(\mathbf{u}_m)\geq D^2,~\forall n\neq m.
\end{align}
 According to \eqref{eq11} and \eqref{eq14}, in the $t$-th iteration of SCA, the optimization problem of the $m$-th MA's position $\mathbf{u}_m$ can be transformed into
 \begin{subequations}
 	\label{eq15}
 	\begin{align}
 		\max_{\mathbf{u}_m}& \quad \Phi^{\mathrm{lb},t}(\mathbf{u}_m) \\
 		\mathrm{s.t.} 
 		 & \quad \mathrm{\eqref{eq10b},\eqref{eq14}},\notag
 	\end{align}
 \end{subequations}
 which is a standard quadratic programming (QP) problem and thus can be efficiently solved via the CVX toolbox\cite{ref32}. Therefore, a suboptimal solution to problem \eqref{eq9} can be obtained by alternately optimizing each MA position, $\mathbf{u}_m,\forall m\in\mathcal{M}$, until the increment of the objective value of \eqref{eq9a} over two consecutive iterations is less than a prescribed value. Finally, denoting the optimized MA positions as $\mathbf{u}_m^{\star},\forall m\in\mathcal{M}$, the SIC decoding order can thus be determined as the increasing order of the corresponding users' channel gains with respect to the optimized $\tilde{\mathbf{u}}^{\star}=\left[\mathbf{u}_1^{\star\mathrm{T}},\dots,\mathbf{u}_M^{\star\mathrm{T}}\right]^{\mathrm{T}}$, i.e., 
 \begin{align}
 	\label{eq16}
 	\left\|\mathbf{h}_{z_1}(\tilde{\mathbf{u}}^{\star})\right\|_2^2\leq\left\|\mathbf{h}_{z_2}(\tilde{\mathbf{u}}^{\star})\right\|_2^2\leq\dots\leq\left\|\mathbf{h}_{z_K}(\tilde{\mathbf{u}}^{\star})\right\|_2^2.
 \end{align}

\begin{figure}[t]
	\vspace{-0.3cm}
	\begin{algorithm}[H]
		\caption{SIC Decoding Order Determination Algorithm}
		\begin{algorithmic}[1]
			\STATE Initialize a feasible solution $\{\mathbf{u}_m^0\}$ to problem \eqref{eq9} and set $t=0$.
			
			\REPEAT
			
			\FOR{$m = 1: M$}
			
			\STATE Compute $\nabla\Phi(\mathbf{u}_m^t)$ and $\delta_{m}$ via \eqref{eq40} and \eqref{eq12}.
			
			\STATE Obtain $\mathbf{u}_m^{t+1}$ by solving problem \eqref{eq15}.
			
			\ENDFOR
			
			\STATE $t\leftarrow t+1$.
			
			\UNTIL{the fractional increase of the objective value is less than a predefined threshold $\epsilon_1>0$.}
			
			\STATE Obtain the SIC decoding order $\left\{z_k\right\}$ according to \eqref{eq16}.
		\end{algorithmic}
		\label{alg_1}
	\end{algorithm}
	\vspace{-0.5cm}
\end{figure}
 The detailed procedure of the proposed SIC decoding order determination algorithm is summarized in Algorithm 1. It can be verified similarly as in \cite{ref4} that the SCA iterations described in lines 3-6 yield a non-decreasing objective value. Meanwhile, the objective in \eqref{eq9a} is inherently upper-bounded based on its definition and the feasible solution set is compact. Thus, the convergence of Algorithm 1 is guaranteed. The computational
 complexity of this algorithm is analyzed as follows. In line 4, the computational complexity for evaluating $\nabla\Phi(\mathbf{u}_m^t)$ and $\delta_{m}$ is $\mathcal{O}\left(KL_k^2\right)$. Besides, in line 5, the computational
 complexity for solving the QP problem \eqref{eq15} is $\mathcal{O}\left(M^{1.5}\ln\frac{1}{\varepsilon}\right)$, where $\varepsilon>0$ denotes the solution accuracy. Furthermore, in line 9, the computational complexity of sorting the users' channel gains is  $\mathcal{O}\left(K\log_2K\right)$. As a result, the overall complexity of Algorithm 1 is $\mathcal{O}\left(I_1M\left(KL_k^2+M^{1.5}\ln\frac{1}{\varepsilon}\right)+K\log_2K\right)$, where $I_1$ denotes the maximum number of iterations required for repeating the procedures in lines 3-7.

\subsection{Joint Design of Beamforming Vectors, MA Positions, and Decoding Indicator Matrix}
In this subsection, given the SIC decoding order (obtained from the first stage), we aim to jointly optimize the transmit beamforming vectors, $\left\{\mathbf{w}_k\right\}$, the APV of all MAs, $\tilde{\mathbf{u}}$, and the decoding indicator matrix, $\mathbf{\Pi}$. For notational simplicity, we reindex each user according to the determined SIC decoding order, i.e., $k\leftarrow z_k,\forall k\in\mathcal{K}$. Then, the optimization problem is transformed into
\begin{subequations}
	\label{eq17}
	\begin{align}
		\max_{\tilde{\mathbf{u}},\{\mathbf{w}_k\},\mathbf{\Pi}}& \quad \sum_{k=1}^{K}R_{k}\label{eq17a}\\
		\mathrm{s.t.}
		& \quad \mathbf{u}_m\in\mathcal{C}_t,~\forall m \in \mathcal{M},\label{eq17b}\\
		& \quad \|\mathbf{u}_m-\mathbf{u}_n\|_2\geq D,~\forall m,n\in\mathcal{M},m\neq n,\label{eq17c}\\
		& \quad \sum_{k=1}^{K}\|\mathbf{w}_k\|^2\leq P_{\max},\label{eq17d}\\
		& \quad R_{k}\geq R_k^{\min},~\forall k \in \mathcal{K},\label{eq17e}\\
		& \quad \pi_{k,i}\in\left\{0,1\right\}, \forall k,i\in\mathcal{K},k< i.\label{eq17f}
	\end{align}
\end{subequations}
Problem \eqref{eq17} is highly non-convex due to the intricate coupling among the optimization variables. To tackle this challenge, we first transform the problem into a more tractable form. By introducing two sets of slack optimization variables, $\{\alpha_{k,i}\}$ and $\{\beta_{k,i}\}$, problem \eqref{eq17} can be equivalently transformed into
\begin{subequations}
	\label{eq18}
	\begin{align}
		\max_{\tilde{\mathbf{u}},\{\mathbf{w}_k,R_{k},\alpha_{k,i},\beta_{k,i}\}}& \quad \sum_{k=1}^{K}R_{k}\\
		\mathrm{s.t.}
		& \quad R_{k}\leq \log_2\left(1+\frac{1}{\alpha_{k,i}\beta_{k,i}}\right),\notag\\
		&\quad ~\forall k,i\in\mathcal{K},~ \pi_{k,i}=1,\label{eq18b}\\
		& \quad \frac{1}{\alpha_{k,i}}\leq \left|\mathbf{h}_i^{\mathrm{H}}\left(\tilde{\mathbf{u}}\right)\mathbf{w}_k\right|^2,~\forall k,i\in\mathcal{K},\label{eq18c}\\
		& \quad \beta_{k,i}\geq\sum\limits_{j=1}^{K}\left|\mathbf{h}_{i}^{\mathrm{H}}\left(\tilde{\mathbf{u}}\right)\mathbf{w}_{j}\right|^2-\sum\limits_{j=1}^{k}\pi_{j,i}\notag\\
		& \quad \times\left|\mathbf{h}_{i}^{\mathrm{H}}\left(\tilde{\mathbf{u}}\right)\mathbf{w}_{j}\right|^2+\sigma_{i}^2,~\forall k,i\in\mathcal{K},\label{eq18d}\\
		& \quad \eqref{eq17b}-\eqref{eq17f}.\notag
	\end{align}
\end{subequations}
Next, we introduce an alternating optimization framework to decompose problem \eqref{eq18} into several subproblems and address them efficiently by exploiting the SCA and GA techniques. 

\textit{1) Optimizing $\{\mathbf{w}_k\}$ with given $\tilde{\mathbf{u}}$ and $\mathbf{\Pi}$:} First, we define $\mathbf{W}_k=\mathbf{w}_k\mathbf{w}_k^{\mathrm{H}},\forall k \in \mathcal{K}$. With any given APV $\tilde{\mathbf{u}}$ and decoding indicator matrix $\mathbf{\Pi}$, problem \eqref{eq18} can be re-expressed as
\begin{subequations}
	\label{eq19}
	\begin{align}
		\max_{\{\mathbf{W}_k,R_{k},\alpha_{k,i},\beta_{k,i}\}}& \quad \sum_{k=1}^{K}R_{k}\\
		\mathrm{s.t.}
		& \quad \frac{1}{\alpha_{k,i}}\leq\mathrm{Tr}\left(\mathbf{W}_k\mathbf{h}_i\mathbf{h}_i^{\mathrm{H}}\right),~\forall k,i\in\mathcal{K},\label{eq19b}\\
		& \quad \beta_{k,i}\geq\sum_{j=1}^{K}\mathrm{Tr}\left(\mathbf{W}_j\mathbf{h}_i\mathbf{h}_i^{\mathrm{H}}\right)-\sum\limits_{j=1}^{k}\pi_{j,i}\notag\\
		& \quad \times\mathrm{Tr}\left(\mathbf{W}_j\mathbf{h}_i\mathbf{h}_i^{\mathrm{H}}\right)+\sigma_i^2,~\forall k,i\in\mathcal{K},\\
		& \quad \sum_{k=1}^{K}\mathrm{Tr}\left(\mathbf{W}_k\right)\leq P_s,\\
		& \quad \mathbf{W}_k\succeq\mathbf{0},~\forall k \in\mathcal{K},\label{eq19e}\\
		& \quad \mathrm{rank}(\mathbf{W}_k)=1,~\forall k \in\mathcal{K},\label{eq19f}\\
		& \quad \eqref{eq18b},\eqref{eq17e}.\notag
	\end{align}
\end{subequations}
Note that problem \eqref{eq19} is still non-convex due to non-convex constraints \eqref{eq19f} and \eqref{eq18b}. Since the right-hand-side (RHS) in constraint \eqref{eq18b} is jointly convex with respect to $\alpha_{k,i}$ and $\beta_{k,i}$, we can utilize SCA to handle it. To this end, by applying the first-order
Taylor series expansion, a lower bound on $\log_2\left(1+\frac{1}{\alpha_{k,i}\beta_{k,i}}\right)$ at any given local points $\left\{\alpha_{k,i}^{t},\beta_{k,i}^{t}\right\}$ can be obtained as
\begin{align}
	\label{eq20}
	\log_2\left(1+\frac{1}{\alpha_{k,i}\beta_{k,i}}\right)&\geq\log_2\left(1+\frac{1}{\alpha_{k,i}^{t}\beta_{k,i}^{t}}\right)\notag\\
	&-\frac{\log_2e}{\alpha_{k,i}^t+{\alpha_{k,i}^t}^2\beta_{k,i}^t}\left(\alpha_{k,i}-\alpha_{k,i}^t\right)\notag\\
	&-\frac{\log_2e}{\beta_{k,i}^t+{\beta_{k,i}^t}^2\alpha_{k,i}^t}\left(\beta_{k,i}-\beta_{k,i}^t\right)\notag\\
	&\triangleq \vartheta_{k,i}^t\left(\alpha_{k,i},\beta_{k,i}\right).
\end{align}
As such, in the $t$-th iteration of SCA, a convex subset of constraint \eqref{eq18b} is established as
\begin{align}
	\label{eq21}
	R_{k}\leq\vartheta_{k,i}^t\left(\alpha_{k,i},\beta_{k,i}\right),~\forall k,i\in\mathcal{K},~ \pi_{k,i}=1,
\end{align}
such that the optimization problem can be recast as
\begin{subequations}
	\label{eq22}
	\begin{align}
		\max_{\{\mathbf{W}_k,R_{k},\alpha_{k,i},\beta_{k,i}\}}& \quad \sum_{k=1}^{K}R_{k}\\
		\mathrm{s.t.}
		& \quad \eqref{eq19b}-\eqref{eq19f},\eqref{eq21},\eqref{eq17e}.\notag
	\end{align}
\end{subequations}
Besides, according to Theorem 1 in \cite{ref38}, the optimal solution of problem \eqref{eq22} without constraint \eqref{eq19f}, i.e., semi-definite relaxation, can always satisfy $\mathrm{rank}(\mathbf{W}_k)=1,\forall k\in\mathcal{K}$. As a result, we can drop the rank-one constraint without loss of its optimality and solve the resulted semidefinite programming (SDP) problem via the CVX toolbox\cite{ref32}. Denoting the optimal solution to problem \eqref{eq22} as $\{\mathbf{W}_k^{\star}\}$, the optimal beamforming vector at the BS, i.e., $\left\{\mathbf{w}_k^{\star}\right\}$, can be obtained from $\{\mathbf{W}_k^{\star}\}$ through eigenvalue decomposition.

\textit{2) Optimizing $\mathbf{u}_m$ with given $\{\mathbf{w}_k\},\mathbf{\Pi}$, and $\{\mathbf{u}_n,n\neq m\}$:} With any given beamforming vectors $\left\{\mathbf{w}_k\right\}$, decoding indicator matrix $\mathbf{\Pi}$, and antenna positions $\left\{\mathbf{u}_n,n\neq m\right\}$, problem \eqref{eq18} can be re-expressed as
\begin{subequations}
	\label{eq23}
	\begin{align}
		\max_{\mathbf{u}_m,\{R_{k},\alpha_{k,i},\beta_{k,i}\}}& \quad \sum_{k=1}^{K}R_{k}\\
		\mathrm{s.t.}
		& \quad \eqref{eq17b},\eqref{eq17c},\eqref{eq17e},\eqref{eq18b}-\eqref{eq18d}.\notag
	\end{align}
\end{subequations}
Problem \eqref{eq23} is a non-convex problem due to non-convex constraints \eqref{eq17c}, \eqref{eq18b}, \eqref{eq18c}, and \eqref{eq18d}, rendering it challenging to acquire the optimal solution. Similarly, we utilize the SCA
technique to address the problem. 

In the previous discussions,
we have shown that \eqref{eq14} and \eqref{eq21} are convex subsets of non-convex constraints \eqref{eq17c} and \eqref{eq18b}, respectively. Next, for handling non-convex constraint \eqref{eq18c}, we first define $\Gamma_{k,i}(\mathbf{u}_m)\triangleq\left|\mathbf{h}_i^{\mathrm{H}}\left(\tilde{\mathbf{u}}\right)\mathbf{w}_k\right|^2,\forall k,i$. Then, denote the gradient and Hessian matrix of $\Gamma_{k,i}(\mathbf{u}_m)$ over $\mathbf{u}_m$ as $\nabla\Gamma_{k,i}(\mathbf{u}_m)$ and $\nabla^2\Gamma_{k,i}(\mathbf{u}_m)$, respectively, with their derivations provided in Appendix \ref{appendix_B}.
Similarly as in \eqref{eq11}, by drawing on Taylor’s theorem,
we can also construct a quadratic concave function to globally lower-bound $\Gamma_{k,i}(\mathbf{u}_m)$, i.e., the RHS of \eqref{eq18c}, denoted as
\begin{align}
	\label{eq24}
	\Gamma_{k,i}\left(\mathbf{u}_m\right)\geq&\Gamma_{k,i}\Big(\mathbf{u}_m^t\Big)+\nabla\Gamma_{k,i}\Big(\mathbf{u}_m^t\Big)^\mathrm{T}\Big(\mathbf{u}_m-\mathbf{u}_m^t\Big)\notag\\
	&-\frac{\gamma_{k,i}}{2}\Big(\mathbf{u}_m-\mathbf{u}_m^t\Big)^{\mathrm{T}}\Big(\mathbf{u}_m-\mathbf{u}_m^t\Big)\notag\\
	\triangleq&\Gamma_{k,i}^{\mathrm{lb},t}\left(\mathbf{u}_m\right),
\end{align}
where $\gamma_{k,i}$ is a positive real number satisfying $\gamma_{k,i}\mathbf{I}_2\succeq\nabla^2\Gamma_{k,i}(\mathbf{u}_m)$. 
Also, since we have $\left\|\nabla^2\Gamma_{k,i}(\mathbf{u}_m)\right\|_{\mathrm{F}}\mathbf{I}_2\succeq\left\|\nabla^2\Gamma_{k,i}(\mathbf{u}_m)\right\|_2\mathbf{I}_2\succeq\nabla^2\Gamma_{k,i}(\mathbf{u}_m)$, the value of $\gamma_{k,i}$ can be determined as
\begin{align}
	\label{eq25}
	&\gamma_{k,i}=\left\|\nabla^2\Gamma_{k,i}(\mathbf{u}_m)\right\|_{\mathrm{F}}=\left[\left(\frac{\partial^2\Gamma_{k,i}(\mathbf{u}_m)}{\partial x_m^2}\right)^2+\right.\\
	&\left.\left(\frac{\partial^2\Gamma_{k,i}(\mathbf{u}_m)}{\partial x_m\partial y_m}\right)^2+\left(\frac{\partial^2\Gamma_{k,i}(\mathbf{u}_m)}{\partial y_m\partial x_m}\right)^2+\left(\frac{\partial^2\Gamma_{k,i}(\mathbf{u}_m)}{\partial y_m^2}\right)^2\right]^{\frac{1}{2}}.\notag
\end{align}
Then, for addressing non-convex constraint \eqref{eq18d}, we define $\Upsilon_{k,i}(\mathbf{u}_m)=\sum_{j=1}^{K}\left|\mathbf{h}_{i}^{\mathrm{H}}\left(\tilde{\mathbf{u}}\right)\mathbf{w}_{j}\right|^2-\sum_{j=1}^{k}\pi_{j,i}\left|\mathbf{h}_{i}^{\mathrm{H}}\left(\tilde{\mathbf{u}}\right)\mathbf{w}_{j}\right|^2+\sigma_{i}^2,\forall k,i$. Similarly, based on Taylor's theorem, we can construct a quadratic convex upper bound on $\Upsilon_{k,i}(\mathbf{u}_m)$, i.e., the RHS of \eqref{eq18d}, denoted as
\begin{align}
	\label{eq26}
	\Upsilon_{k,i}\left(\mathbf{u}_m\right)\leq&\Upsilon_{k,i}\Big(\mathbf{u}_m^t\Big)+\nabla\Upsilon_{k,i}\Big(\mathbf{u}_m^t\Big)^\mathrm{T}\Big(\mathbf{u}_m-\mathbf{u}_m^t\Big)\notag\\
	&+\frac{\psi_{k,i}}{2}\Big(\mathbf{u}_m-\mathbf{u}_m^t\Big)^{\mathrm{T}}\Big(\mathbf{u}_m-\mathbf{u}_m^t\Big)\notag\\
	\triangleq&\Upsilon_{k,i}^{\mathrm{ub},t}\left(\mathbf{u}_m\right),
\end{align}
where the gradient of $\Upsilon_{k,i}(\mathbf{u}_m)$ with respect to $\mathbf{u}_m$ and the positive real number $\psi_{k,i}$ satisfying $\psi_{k,i}\mathbf{I}_2\succeq\nabla^2\Upsilon_{k,i}(\mathbf{u}_m)$ can be respectively calculated as
\begin{align}
	&\nabla\Upsilon_{k,i}\left(\mathbf{u}_m\right)=\sum_{j=1}^{K}\nabla\Gamma_{j,i}\left(\mathbf{u}_m\right)-\sum_{j=1}^{k}\pi_{j,i}\nabla\Gamma_{j,i}\left(\mathbf{u}_m\right),\label{eq27}\\
	&\psi_{k,i}=\sum_{j=k+1}^{K}\gamma_{j,i}+\sum_{j=1}^{k}(1-\pi_{j,i})\gamma_{j,i}.\label{eq28}
\end{align}

Armed with \eqref{eq24} and \eqref{eq26}, in the $t$-th iteration of SCA, the optimization problem of the $m$-th MA's position $\mathbf{u}_m$ can be transformed into
\begin{subequations}
	\label{eq29}
	\begin{align}
		\max_{\mathbf{u}_m,\{R_{k},\alpha_{k,i},\beta_{k,i}\}}& \quad \sum_{k=1}^{K}R_{k}\\
		\mathrm{s.t.} & \quad \frac{1}{\alpha_{k,i}}\leq\Gamma_{k,i}^{\mathrm{lb},t}\left(\mathbf{u}_m\right),~\forall k,i\in\mathcal{K},\\
		& \quad \beta_{k,i}\geq \Upsilon_{k,i}^{\mathrm{ub},t}\left(\mathbf{u}_m\right),~\forall k,i\in\mathcal{K},\\
		& \quad \eqref{eq17b},\eqref{eq17e},\eqref{eq14},\eqref{eq21},\notag
	\end{align}
\end{subequations}
which is a convex problem and thus can be efficiently solved via the CVX toolbox\cite{ref32}. 

\textit{3) Optimizing $\mathbf{\Pi}$ with given $\{\mathbf{w}_k\}$ and $\tilde{\mathbf{u}}$:} With any given transmit beamforming vectors $\{\mathbf{w}_k\}$ and APV $\tilde{\mathbf{u}}$, the optimization problem can be re-expressed as
\begin{subequations}
	\label{eq30}
	\begin{align}
		\max_{\mathbf{\Pi}}& \quad \sum_{k=1}^{K}R_{k}\\
		\mathrm{s.t.}
		& \quad \eqref{eq17e},\eqref{eq17f}.\notag
	\end{align}
\end{subequations}
which is a combinatorial optimization problem with respect to the elements in $\mathbf{\Pi}$. To address this non-convex problem, we adopt a low-complexity heuristic method based on GA to acquire a suboptimal solution\cite{ref68}. Specifically, we first randomly initialize $G$ individuals with genes $\{\mathcal{G}_{g}^{(0)}\}_{g=1}^{G}$, where each individual's gene $\mathcal{G}_{g}^{(0)}$ denotes a truncated expression of a possible solution to problem \eqref{eq30}, i.e.,
\begin{align}
	\label{eq31}
	\mathcal{G}_{g}^{(0)}=\left[\pi_{1,2}^{g,(0)},\cdots,\pi_{1,K}^{g,(0)},\pi_{2,3}^{g,(0)},\cdots,\pi_{2,K}^{g,(0)},\cdots,\pi_{K-1,K}^{g,(0)}\right]^{\mathrm{T}}.
\end{align}
For each individual, the fitness function is defined as 
\begin{align}
	\label{eq32}
	\mathcal{F}\left(\mathcal{G}_{g}^{(v)}\right)=\sum_{k=1}^{K}R_k\left(\mathcal{G}_{g}^{(v)}\big|\{\mathbf{w}_k\},\tilde{\mathbf{u}}\right)-\tau\left|\mathcal{P}\left(\mathcal{G}_{g}^{(v)}\right)\right|,
\end{align}
where $v$ is the generation index of GA and $R_k\left(\mathcal{G}_{g}^{(v)}\big|\{\mathbf{w}_k\},\tilde{\mathbf{u}}\right)$ can be calculated by \eqref{eq7} with given $\{\mathbf{w}_k\}$ and $\tilde{\mathbf{u}}$. Besides, in order to ensure constraint \eqref{eq17e}, we introduce a penalty factor $\tau\left|\mathcal{P}\left(\mathcal{G}_{g}^{(v)}\right)\right|$, where 
$\mathcal{P}\left(\mathcal{G}_{g}^{(v)}\right)$ is a set defined as $\mathcal{P}\left(\mathcal{G}_{g}^{(v)}\right)\triangleq\left\{R_k|R_k\leq R_k^{\min}, \forall k \in\mathcal{K}\right\}$, and $\tau$ is a large positive penalty parameter such that $\tau\geq\sum_{k=1}^{K}R_k$ holds for all individuals.

In each generation, we employ the roulette wheel selection method to choose $G$ parent individuals for crossover. Particularly, the probability for each individual to be selected is
\begin{align}
	\label{eq33}
	P\left(\mathcal{G}_{g}^{(v)}\right)=\frac{\mathcal{F}\left(\mathcal{G}_{g}^{(v)}\right)}{\sum_{j=1}^{G}\mathcal{F}\left(\mathcal{G}_{j}^{(v)}\right)},
\end{align}
which indicates that the individuals with higher fitness values are more likely to be selected. Subsequently, the $G$ parent individuals are randomly divided into $G/2$ groups, with two individuals in each group performing crossover to generate their offspring. Specifically, denoting $\{\mathcal{G}_{i}^{(v)},\mathcal{G}_{j}^{(v)}\}$ as a parent pair in the $v$-th generation, their two offspring individuals can be generated as
\begin{subequations}
	\label{eq34}
	\begin{align}
		\mathcal{G}_{i,\mathrm{child}}^{(v)}=\left[\mathcal{C}(\pi_{1,2}^{i,(v)},\pi_{1,2}^{j,(v)}),\cdots,\mathcal{C}(\pi_{K-1,K}^{i,(v)},\pi_{K-1,K}^{j,(v)})\right]^{\mathrm{T}},\\
		\mathcal{G}_{j,\mathrm{child}}^{(v)}=\left[\mathcal{C}(\pi_{1,2}^{j,(v)},\pi_{1,2}^{i,(v)}),\cdots,\mathcal{C}(\pi_{K-1,K}^{i,(v)},\pi_{K-1,K}^{j,(v)})\right]^{\mathrm{T}},
	\end{align}
\end{subequations}
where $\mathcal{C}(\cdot)$ is the crossover function defined as
\begin{align}
	\label{eq35}
	\mathcal{C}(\pi_{a,b}^i,\pi_{a,b}^j)=\begin{cases}
		\pi_{a,b}^j & \text{, if $c_{a,b}\leq p_{\mathrm{c}}$,}\\
		\pi_{a,b}^i & \text{, otherwise.}
	\end{cases}
\end{align}
In \eqref{eq35}, $c_{a,b}$ denotes a random parameter with uniform distribution over $[0,1]$ and $p_{\mathrm{c}}$ is the crossover probability. For each offspring, if its fitness value is lower than that of its parents, it will be directly eliminated while its parent will be inherited into the next generation instead, i.e.,
\begin{align}
	\label{eq36}
	\mathcal{G}_{g,\mathrm{next}}^{(v)}=\arg\max\limits_{\mathcal{G}}\left(\mathcal{F}\left(\mathcal{G}_{g,\mathrm{child}}^{(v)}\right),\mathcal{F}\left(\mathcal{G}_{g}^{(v)}\right)\right).
\end{align}

\begin{figure}[t]
	\vspace{-0.3cm}
	\begin{algorithm}[H]
		\caption{GA-Based Approach for Solving Problem \eqref{eq30}}
		\begin{algorithmic}[1]
			\STATE Initialize $G$ individuals with genes $\{\mathcal{G}_{g}^{(0)}\}_{g=1}^{G}$ and the best individual $\mathcal{G}_{\mathrm{best}}=\arg\max\left(\mathcal{F}(\mathcal{G}_g^{(0)})\right)$; set $v=0$.
			
			\REPEAT
			
			\STATE Select the parent individuals from $\{\mathcal{G}_g^{(v)}\}$ by using the roulette method.
			
			\STATE Divide the parents into $G/2$ groups for crossover and generate their offspring $\{\mathcal{G}_{g,\mathrm{child}}^{(v)}\}$ via \eqref{eq34}.
			
			\STATE Update the candidate individuals $\{\mathcal{G}_{g,\mathrm{next}}^{(v)}\}$ via \eqref{eq36}.
			
			\STATE Perform the mutation operation for each individual and update the candidate individuals $\{\mathcal{G}_{g,\mathrm{next}'}^{(v)}\}$ via \eqref{eq37}.
			
			\STATE Update the individual set as $\{\mathcal{G}_g^{(v+1)}\}=\{\mathcal{G}_{g,\mathrm{next}'}^{(v)}\}$ and compute each individual's fitness value via \eqref{eq32}.
			
			\STATE Update the best individual according to $\mathcal{G}_{\mathrm{best}}=\arg\max\left(\mathcal{F}(\mathcal{G}_{\mathrm{best}}),\{\mathcal{F}(\mathcal{G}_g^{(v+1)})\}\right)$.
			
			\STATE $v\leftarrow v+1$.
			
			\UNTIL the maximum generation index $V_{\max}$ is reached.
			
			\STATE Obtain the best individual $\mathcal{G}_{\mathrm{best}}$ and reconstruct the decoding indicator matrix $\mathbf{\Pi}$ from $\mathcal{G}_{\mathrm{best}}$.
		\end{algorithmic}
		\label{alg_2} 
	\end{algorithm}
	\vspace{-0.5cm}
\end{figure}
Besides, to avoid plunging into the undesired local optima, in each generation, a mutation operation is carried out for all individuals. Let $p_{\mathrm{m}}$ denote the mutation probability. The new individual $\mathcal{G}_{g,\mathrm{next}'}^{(v)}$ mutated from $\mathcal{G}_{g,\mathrm{next}}^{(v)}$ can be expressed as
\begin{align}
	\label{eq37}
	\mathcal{G}_{g,\mathrm{next}'}^{(v)}=\left[\pi_{1,2}^{g,\mathrm{next},(v)},\cdots,\tilde{\pi}_{i,j}^{g,\mathrm{next},(v)},\cdots,\pi_{K-1,K}^{g,\mathrm{next},(v)}\right]^{\mathrm{T}},
\end{align}
where the element $\pi_{i,j}^{g,\mathrm{next},(v)}$ is randomly selected from $\mathcal{G}_{g,\mathrm{next}}^{(v)}$ and replaced by
\begin{align}
	\label{eq38}
	\tilde{\pi}_{i,j}^{g,\mathrm{next},(v)}=\begin{cases}
		1-\pi_{i,j}^{g,\mathrm{next},(v)} & \text{, if $d_g\leq p_{\mathrm{m}}$,}\\
		\pi_{i,j}^{g,\mathrm{next},(v)} & \text{, otherwise,}
	\end{cases}
\end{align}
with $d_g$ being a random parameter uniformly distributed over $[0,1]$.
Finally, the whole individual set inherited into the next generation is updated as $\{\mathcal{G}_g^{(v+1)}\}=\{\mathcal{G}_{g,\mathrm{next}'}^{(v)}\}$. 

The details of the proposed GA-based approach for solving problem \eqref{eq30} are summarized in Algorithm 2. In each generation, the parent selection procedure corresponds to the step in line 3; the crossover operation corresponds to the step in lines 4-5; the mutation operation corresponds to the step in lines 6-7. In line 8, the best individual $\mathcal{G}_{\mathrm{best}}$ within all generations is updated. Finally, the iteration of GA is terminated until the maximum generation index $V_{\max}$ is reached and a suboptimal solution to problem \eqref{eq30} is reconstructed from $\mathcal{G}_{\mathrm{best}}$.



\subsection{Overall Two-Stage Algorithm}
\begin{figure}[t]
	\vspace{-0.3cm}
	\begin{algorithm}[H]
		\caption{Overall Two-Stage Algorithm for Solving Problem \eqref{eq8}}
		\begin{algorithmic}[1]
			\STATE Initialize a feasible solution $\left\{\{\mathbf{w}_k^0\},\{\mathbf{u}_m^0\},\mathbf{\Pi}^0\right\}$ and set $t=0$.
			
			\STATE \textbf{Stage one:} SIC decoding order determination
			
			\STATE Obtain the SIC decoding order $\left\{z_k\right\}$ by applying Algorithm 1.
			
			\STATE \textbf{Stage two:} Joint design of transmit beamforming vectors, MA positions, and decoding indicator matrix
			
			\REPEAT
			
			\STATE Obtain $\left\{\mathbf{w}_k^{t+1}\right\}$ by solving problem \eqref{eq22}.
			
			\FOR{$m = 1: M$}
			\STATE Compute $\left\{\nabla\Gamma_{k,i}(\mathbf{u}_m^t),\gamma_{k,i},\nabla\Upsilon_{k,i}(\mathbf{u}_m^t),\psi_{k,i}\right\}$ via \eqref{eq43}, \eqref{eq25}, \eqref{eq27}, and \eqref{eq28}, respectively.
			\STATE Obtain $\mathbf{u}_m^{t+1}$ by solving problem \eqref{eq29}. 
			\ENDFOR
			
			\STATE Obtain $\mathbf{\Pi}'$ according to Algorithm 2.
			
			\IF{$\sum_{k=1}^{K}R_k\left(\mathbf{\Pi}'\right)>\sum_{k=1}^{K}R_k\left(\mathbf{\Pi}^{t}\right)$}
			\STATE Obtain $\mathbf{\Pi}^{t+1}=\mathbf{\Pi}'$.
			\ELSE
			\STATE Obtain $\mathbf{\Pi}^{t+1}=\mathbf{\Pi}^{t}$.
			\ENDIF
			
			\STATE $t\leftarrow t+1$.
			
			\UNTIL the fractional increase of the objective value is less than a predefined threshold $\epsilon_2>0$.
			
			\STATE \textbf{return} $\left\{z_k\right\},\left\{\mathbf{w}_k\right\},\left\{\mathbf{u}_m\right\}$, and $\mathbf{\Pi}$.
		\end{algorithmic}
		\label{alg_3} 
	\end{algorithm}
	\vspace{-0.5cm}
\end{figure}
The overall two-stage algorithm for addressing problem \eqref{eq8} is summarized in Algorithm 3. Lines 2-3 correspond to the first stage of our proposed algorithm, where the SIC decoding order is determined by applying Algorithm 1. Lines 4-12 correspond to the second stage of our proposed algorithm, where the beamforming vectors, MA positions, and decoding indicator matrix are iteratively optimized in an alternating optimization framework. Note that since the objective is non-decreasing over the iterations and the solution set is compact, the convergence of the optimization procedure in stage two is guaranteed. 

Besides, the computational
complexity of this algorithm is analyzed as follows. In line 6, the computational complexity for solving problem \eqref{eq22} is $\mathcal{O}\Big(\max(M,K^2+K)^4M^{1/2}\ln\frac{1}{\varepsilon}\Big)$. 
The computational complexities of lines 8-9 are $\mathcal{O}\big(K^2L_k^2\big)$ and $\mathcal{O}\Big((K^2+M)^{1.5}\ln\frac{1}{\varepsilon}\Big)$, respectively. The computational complexity of line 11 is $\mathcal{O}\Big(V_{\max}\big(G+p_{\mathrm{c}}K(K-1)G/2+p_{\mathrm{m}}G\big)\Big)$, which is based on the complexities of the parent selection, crossover operation, and mutation operation in Algorithm 2. Therefore, the overall computational complexity of our proposed two-stage algorithm is $\mathcal{O}\Big(I_1M\big(KL_k^2+M^{1.5}\ln\frac{1}{\varepsilon}\big)+K\log_2K\Big)+\mathcal{O}\Big(I_2\max(M,K^2+K)^4M^{1/2}\ln\frac{1}{\varepsilon}\Big)+\mathcal{O}\Big(I_2M\big(K^2L_k^2+(K^2+M)^{1.5}\ln\frac{1}{\varepsilon}\big)\Big)+\mathcal{O}\Big(I_2V_{\max}\big(G+p_{\mathrm{c}}K(K-1)G/2+p_{\mathrm{m}}G\big)\Big)$, where $I_2$ denotes the maximum number of iterations for repeating lines 6-17 in stage two.

\section{Simulation Results}\label{section_4}
In this section, simulation results are provided to evaluate the performance of our proposed MA array-enhanced downlink NOMA system and validate the effectiveness of the proposed algorithm for maximizing the users' sum rate.

\subsection{Simulation Setup}
In our simulations, we employ the geometric channel model\cite{ref3}, in which the numbers of  transmit channel paths from the BS to all users are identical, i.e., $L_k=L=5,\forall k\in\mathcal{K}$. The PRV for each user is modeled as a circularly symmetric complex Gaussian (CSCG) random vector with each element satisfying $f_{k,\ell}\sim\mathcal{CN}\left(0,\rho d_k^{-\alpha}/L\right),1\leq \ell\leq L$, where $\rho=-30$ dB denotes the expected path loss at the reference distance, i.e., 1 meter (m), $d_k$ denotes the distance from the BS to user $k$, and $\alpha=2.8$ represents the path loss exponent. The elevation and azimuth AoDs of the transmit channel paths from the BS to each user are presumed to be independent and identically distributed (i.i.d.) variables, which follow uniform distributions over $[0,\pi]$, i.e., $\theta_{k,\ell},\phi_{k,\ell}\sim\mathcal{U}[0,\pi],1\leq k \leq K,1\leq \ell \leq L$. The distance between the BS and each user are randomly generated from $50$ m to $100$ m. The minimum rate requirement for each user is set identically, i.e., $R_k^{\min}=R_{\min}=0.25~\text{bps/Hz},1\leq k \leq K$; the average noise power for all users is set as $\sigma_k^2=-80$ dBm; the side length of the MA moving region is set as $A=3\lambda$ and the minimum inter-MA distance is set as $D=\lambda/2$\cite{ref9}. Moreover, for our proposed Algorithms 1 and 3, the convergence thresholds are set as $\epsilon_1=\epsilon_2=10^{-2}$. For our proposed Algorithm 2, we set $G=100,\tau=100,p_{\mathrm{c}}=0.5,p_{\mathrm{m}}=0.1$, and $V_{\max}=200$. Besides, all points in our simulation curves are averaged over 1000 independent channel realizations.

\subsection{Convergence of Proposed Algorithms}
\begin{figure}
	\vspace{-0.3cm}
	\centering
	\subfloat[]
	{\includegraphics[width=0.24\textwidth]{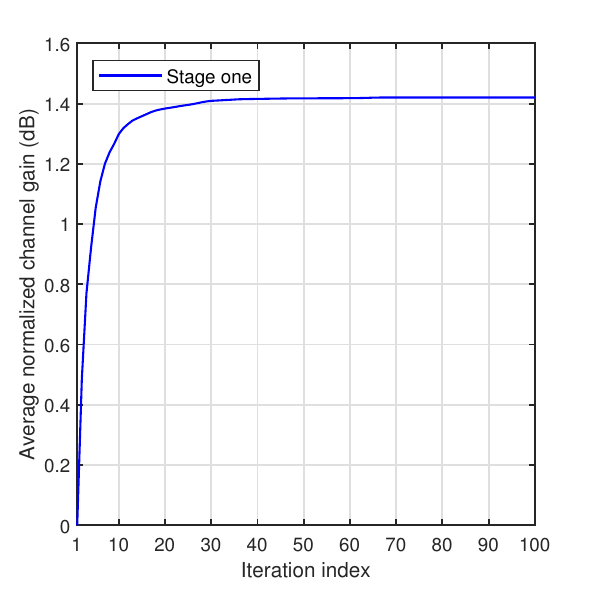}
	\label{convergence_performance_1}}
	\subfloat[]
	{\includegraphics[width=0.24\textwidth]{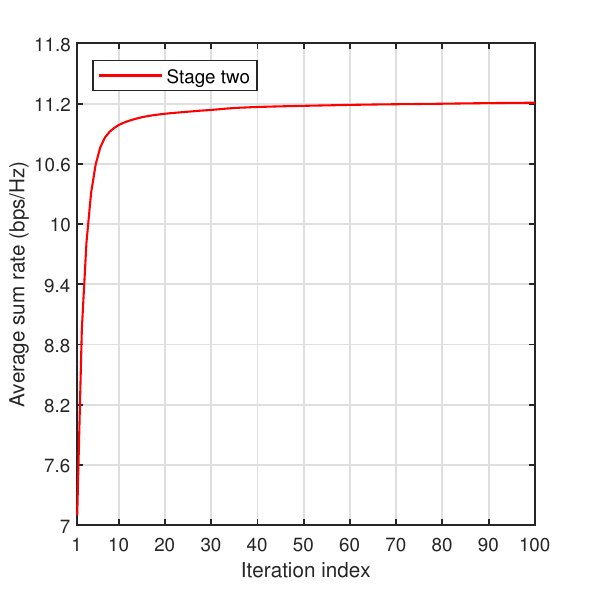}
	\label{convergence_performance_2}}
	\caption{Evaluation of the convergence performance of our proposed two-stage optimization algorithm: (a) Stage one; (b) Stage two.}
	\label{convergence_performance}
	\vspace{-0.2cm}
\end{figure}
In Fig. \ref{convergence_performance}, we illustrate the convergence performance of our proposed Algorithm 3, where the number of antennas at the BS is set as $M=4$, the number of users is set as $K=6$, and the maximum transmit power of the BS is set as $P_{\max}=10~\text{dBm}$. From Fig. \ref{convergence_performance}(a), it can be observed that the normalized channel gain obtained by our proposed algorithm in stage one increases monotonically with the iteration index and remains unchanged after 40 iterations on average. Meanwhile, in Fig. \ref{convergence_performance}(b), we also observe that the average sum rate obtained by our proposed algorithm in stage two increases with the iteration index and converges to a constant after about 60 iterations on average. Thus, combining the results in Figs. \ref{convergence_performance}(a) and \ref{convergence_performance}(b), the convergence of our proposed two-stage optimization algorithm is verified. Besides, the average sum rate in Fig. \ref{convergence_performance}(b) increases from 5.85 bps/Hz to 10.60 bps/Hz, which demonstrates the effectiveness of our proposed algorithm in improving the system's sum rate.

\subsection{Comparison with Benchmark Schemes}
In this subsection, to highlight the great benefits provided by MAs, we compare the proposed MA-enhanced NOMA system, labeled by ``\textbf{NOMA-MA}'', with the following three benchmark schemes.
\begin{itemize}
	\item \textbf{NOMA-FPA}: The BS is equipped with an FAP-based uniform planar array with $M$ antennas, and the $K$ single-antenna users are served by NOMA.
	
	 \item \textbf{SDMA-MA}: The BS is still equipped with an MA array with $M$ antennas, while the $K$ single-antenna users are served by SDMA\cite{ref9}.
	 
	 \item \textbf{SDMA-FPA}: The BS is equipped with an FAP-based uniform planar array with $M$ antennas, and the $K$ single-antenna users are served by SDMA.
\end{itemize}

\begin{figure}[!t]
	\vspace{-0.2cm}
	\centering
	\includegraphics[width=0.43\textwidth]{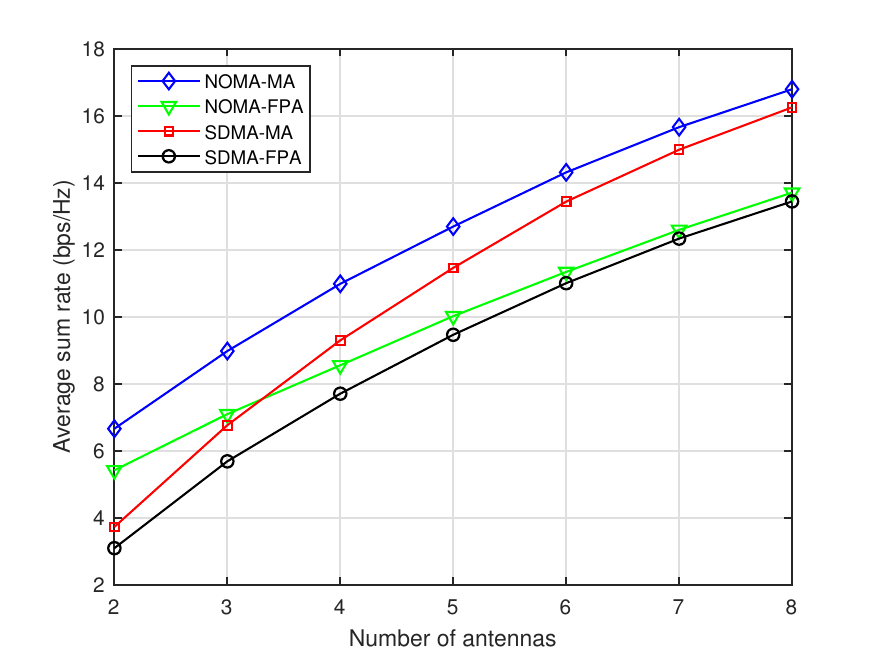}
	\caption{Average sum rate of different schemes versus the number of antennas at the BS for $K=6$ and $P_{\max}=10$ dBm.}
	\label{sum_rate_vs_M}
	\vspace{-0.2cm}
\end{figure}
Fig. \ref{sum_rate_vs_M} presents the average sum rate of different schemes versus the number of antennas at the BS, $M$, where the number of users is set as $K=6$, and the maximum transmit power of the BS is set as $P_{\max}=10$ dBm. As can be observed, as $M$ increases, all considered schemes can achieve a higher average sum rate thanks to the enhanced beamforming gains. Furthermore, the proposed scheme outperforms other benchmark schemes with any value of $M$. Particularly, when $M=4$, our proposed algorithm can achieve 18.2\%, 28.6\%, and 42.6\% performance improvements over the SDMA-MA, NOMA-FPA, SDMA-FPA schemes, respectively. We also observe that for both MA schemes and FPA schemes, the NOMA-based scheme achieves a higher average sum rate compared to its SDMA counterpart, since it allows each user to eliminate the interference from other users, thereby improving the spectral efficiency and system throughput. However, as $M$ grows, the performance gain between NOMA and SDMA diminishes. The reason is that if the number of antennas at the BS is sufficiently large $(M> K)$, the channel correlation among multiple users is relatively small such that the SDMA-based scheme is able to synthesize very sharp energy-focused beams towards each user with low inter-user interference. In such a condition, the NOMA-based schemes will achieve a comparable performance to SDMA-based schemes, since the SIC operation of NOMA can only provide a slight SINR improvement for each user due to the low-level co-channel interference. Lastly, it is interesting to note that the performance gain of the proposed NOMA-MA scheme over the SDMA-MA scheme is larger than that of the NOMA-FPA scheme over the SDMA-FPA scheme. This is because integrating MA into NOMA systems can not only improve channel power gain for each user but also magnify their channel disparities, thereby further enhancing the rate advantage of NOMA over conventional SDMA.

\begin{figure}[!t]
	\centering
	\includegraphics[width=0.43\textwidth]{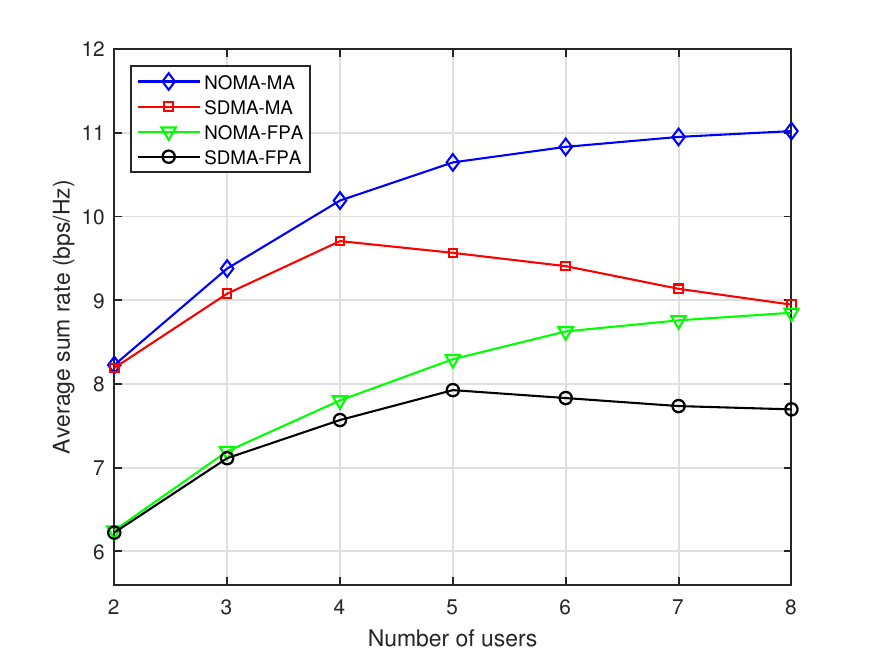}
	\caption{Average sum rate of different schemes versus the number of users for $M=4$ and $P_{\max}=10$ dBm.}
	\label{sum_rate_vs_K}
	\vspace{-0.2cm}
\end{figure}
Fig. \ref{sum_rate_vs_K} illustrates the average sum rate of different schemes versus the number of users, $K$, where the number of antennas at the BS is set as $M=4$, and the maximum transmit power of the BS is set as $P_{\max}=10$ dBm. We can observe that the average sum rates of both NOMA-MA and NOMA-FPA schemes increase as the number of users grows due to the multiplexing gain in power domain. However, the average sum rates of their SDMA counterparts decrease when $K>4$ (the number of antennas). This because SDMA only serves each user via a single beam while NOMA can further utilize SIC to eliminate the interference among multiple users. Especially, when $K>M$, the inter-user interference becomes more pronounced and will result in performance degradation for the SDMA schemes. This phenomenon validates the rate advantage of NOMA in multi-user communications. Besides, we can also find that our proposed algorithm achieves the highest average sum rate among all considered schemes under any value of $K$, which again demonstrates the superiority of our proposed MA-enhanced NOMA system. Similarly, we note that when $K<M$, the performance gap between the NOMA schemes and SDMA schemes is moderate, which is consistent with the observation from Fig. \ref{sum_rate_vs_M}.

\begin{figure}[!t]
	\centering
	\includegraphics[width=0.43\textwidth]{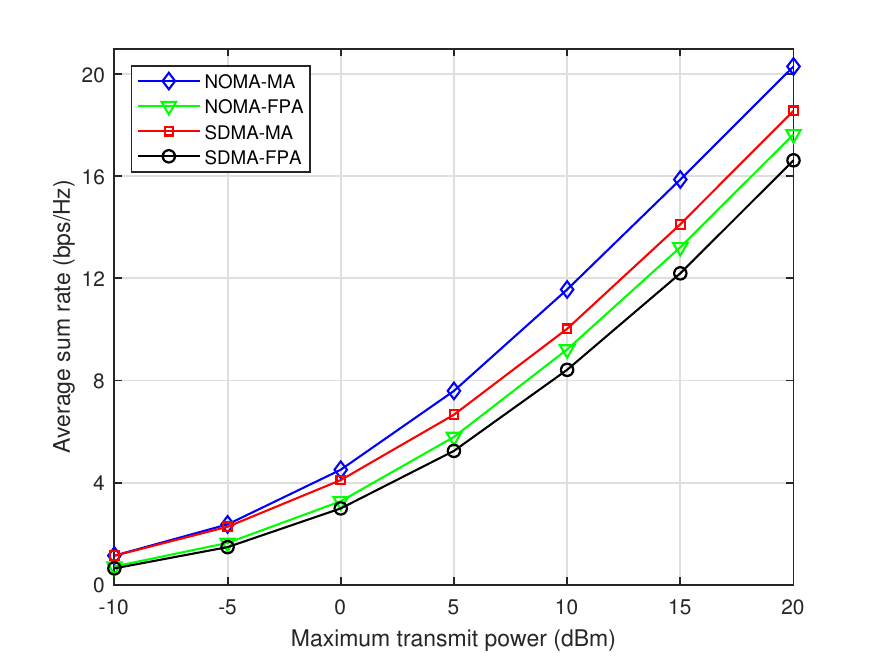}
	\caption{Average sum rate of different schemes versus the maximum transmit power for $M=4$ and $K=6$.}
	\label{sum_rate_vs_Ps}
	\vspace{-0.2cm}
\end{figure}
In Fig. \ref{sum_rate_vs_Ps}, we plot the average sum rate of different schemes versus the maximum transmit power at the BS, $P_{\max}$, where the number of antennas at the BS is set as $M=4$, and the number of users is set as $K=6$. It is first shown that the sum rate of all considered  schemes increases with the maximum transmit power. This is expected because the SINR of each user can be improved by effectively exploiting the increased transmit power budget. Besides, result reveals that to achieve the same sum rate, our proposed scheme requires less transmit power than other benchmark schemes. For instance, to obtain an average sum rate of 16 bps/Hz, our proposed NOMA-MA scheme requires about 15 dBm transmit power, while the NOMA-FPA scheme, SDMA-MA scheme, and SDMA-FPA scheme need about 17 dBm, 18 dBm, and 19 dBm, respectively. This verifies the great superiority of combing MA with NOMA over conventional FPA and SDMA systems.

\subsection{Impact of SIC Decoding Order}
In this subsection, we evaluate the impact of the decoding order on the system's sum rate. For this purpose, we compare the following schemes: 1) ``NOMA-MA-Proposed Order'' denotes the proposed SIC decoding order determination algorithm, i.e., Algorithm 1, where the decoding order is obtained by solving a channel gain maximization problem for all users. 2) ``NOMA-MA-Exhaustive Order'' denotes the exhaustive search scheme, where the optimal decoding order is obtained by enumerating all possible solutions. Specifically, for each decoding order, we first derive its corresponding sum rate by solving problem \eqref{eq8}. Then, we can obtain the optimal decoding order by selecting the one that yields the maximum sum rate. 3) ``NOMA-MA-Random Order'' denotes the scheme where the decoding order is obtained by randomly selected one among all possible decoding orders.

\begin{figure}[!t]
	\centering
	\includegraphics[width=0.43\textwidth]{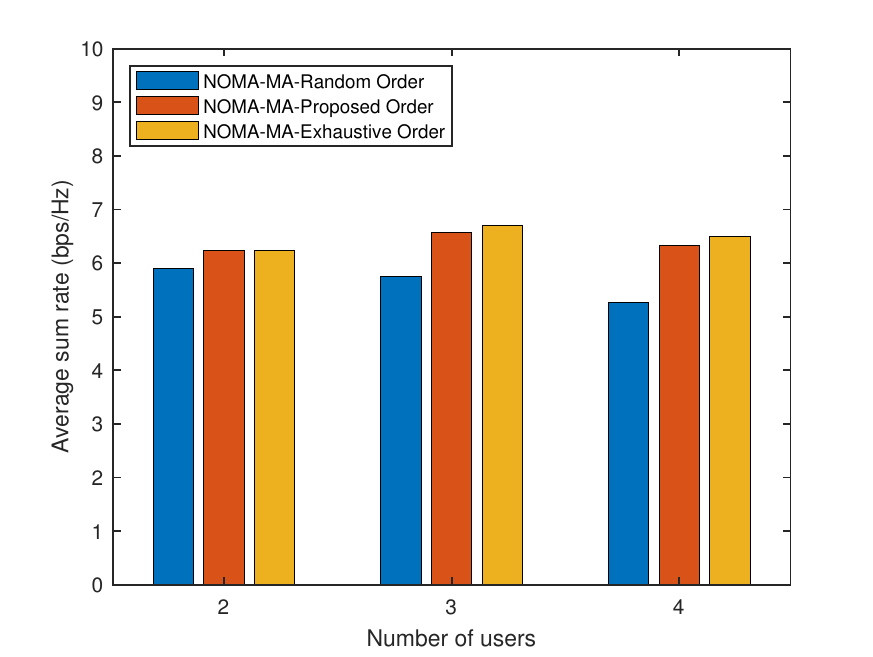}
	\caption{Average sum rate of different decoding orders versus the number of users for $M=2$ and $P_{\max}=10$ dBm.}
	\label{impact_of_decoding_order}
\end{figure}
Fig. \ref{impact_of_decoding_order} shows the average sum rate of different decoding orders versus the number of users, $K$, where the number of antennas at the BS is set as $M=2$ and the maximum transmit power of the BS is set as $P_{\max}=10$ dBm. It is shown that there is a significant performance gap between the ``NOMA-MA-Random Order'' scheme and other two decoding order schemes, demonstrating the importance of determining an appropriate SIC decoding order for MA-enhanced NOMA systems. Besides, it is observed that with any value of $K$, the performance of our proposed scheme can closely approach that of the ``NOMA-MA-Exhaustive Order'' scheme, which demonstrate the effectiveness of our proposed SIC decoding order determination scheme in Algorithm 1.

\subsection{Impact of Adaptive User Decoding}
Finally, we evaluate the impact of users' adaptive decoding scheme on the performance of the considered system. To this end, we introduce the following schemes for comparison: 1) ``NOMA-MA-Proposed Indicator'' denotes the proposed adaptive user decoding scheme, where the decoding indicator matrix $\mathbf{\Pi}$ is optimized by employing the proposed GA-based approach, i.e., Algorithm 2. 2) ``NOMA-MA-Exhaustive Indicator'' denotes the exhaustive search scheme, where the optimal decoding indicator matrix is obtained by enumerating all possible solutions of $\mathbf{\Pi}$. Specifically, for each candidate solution of $\mathbf{\Pi}$, we derive a  sum rate by solving problem \eqref{eq8}. Then, we can obtain the optimal solution by selecting the one yielding the maximum sum rate. 3) ``NOMA-MA-Fixed Indicator'' denotes the conventional fixed NOMA decoding scheme, where each user needs to decode signals for other users with prior decoding users. Note that for the ``NOMA-MA-Fixed Indicator'' scheme, the overall system performance may be significantly deteriorated in some cases such that the minimum rate requirement for each user cannot be satisfied. Therefore, in our simulations, we relax the minimum rate constraints for each user, i.e., $R_k^{\min}=0,\forall k\in\mathcal{K}$.

\begin{figure}[!t]
	\centering
	\includegraphics[width=0.43\textwidth]{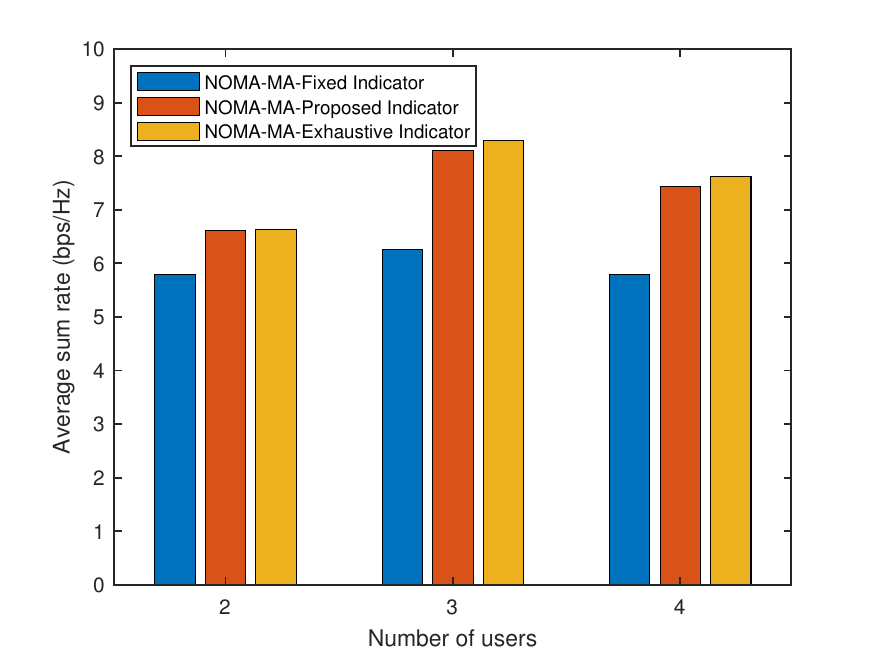}
	\caption{Average sum rate of different decoding indicators versus the number of users for $M=2$ and $P_{\max}=10$ dBm.}
	\label{impact_of_adaptive_decoding}
\end{figure}
In Fig. \ref{impact_of_adaptive_decoding}, we present the average sum rate of different decoding indicators versus the number of users, $K$, where the number of antennas at the BS is set as $M=2$ and the maximum transmit power of the BS is set as $P_{\max}=10$ dBm. As can be observed, the traditional ``NOMA-MA-Fixed Indicator'' scheme suffers from significant performance degradation compared with other two adaptive decoding schemes. This is because the ``NOMA-MA-Fixed Indicator'' scheme enforces each user to decode the signals for all other users with prior decoding orders and does not take into account the channel correlations among multiple users. In contrast, adaptive decoding schemes enable each user to flexibly choose whether to decode signals for other users according to their specific channel conditions, thus providing more DoFs for enhancing system's performance. Furthermore, it is observed that our proposed GA-based adaptive decoding scheme achieves performance close to its exhaustive search counterpart, verifying the effectiveness of Algorithm 2.

\section{Conclusion}\label{section_5}
In this paper, we studied the sum-rate maximization problem for an MA array-enhanced downlink NOMA system, where multiple MAs are deployed at the BS to serve multiple single-antenna users. Specifically, the transmit beamforming vectors, the positions of MAs, the SIC decoding order, and the users' decoding indicator matrix were jointly designed under the maximum BS transmit power and finite MA moving region constraints. To address the non-convex nature of the considered problem with intricate coupling variables, we proposed a two-stage optimization algorithm. In the first stage, we derived the SIC decoding order by solving the channel gain maximization problem. Next, in the second stage, we alternately optimized the beamforming vectors, each MA position, and the decoding indicator matrix by exploiting SCA and GA techniques. Finally, it was shown via simulation results that our proposed MA-aided NOMA system can obtain substantial improvements over conventional FPA-based systems and other benchmark schemes. Besides, our results also unveiled that deploying MAs can further enhance the advantage of NOMA over SDMA. Furthermore, the SIC decoding scheme in our proposed algorithm can achieve a close performance to the exhaustive search method with a low computational complexity, which provides a cost-effective solution for MA-enhanced downlink NOMA systems in practice.

{\appendices
	\section{Derivations of $\nabla\Phi(\mathbf{u}_m)$ and $\nabla^2\Phi(\mathbf{u}_m)$}
	\label{appendix_A}
	To facilitate the expression, let us define $\mathbf{A}_k\triangleq\mathbf{f}_k\mathbf{f}_k^{\mathrm{H}}$, based on which $\Phi(\mathbf{u}_m)$ can be further expanded as
	\begin{align}
		\label{eq39}
		&\Phi(\mathbf{u}_m)=\mathbf{g}_k^{\mathrm{H}}(\mathbf{u}_m)\mathbf{f}_k\mathbf{f}_k^{\mathrm{H}}\mathbf{g}_k(\mathbf{u}_m)=\sum_{k=1}^{K}\mathbf{g}_k^{\mathrm{H}}(\mathbf{u}_m)\mathbf{A}_k\mathbf{g}_k(\mathbf{u}_m)\notag\\
		&=\sum_{k=1}^{K}\sum_{\ell_1=1}^{L_k}\sum_{\ell_2=1}^{L_k}\left[\mathbf{A}_k\right]_{\ell_1,\ell_2}e^{\mathrm{j}\left(\frac{2\pi}{\lambda}\left(-\rho_{k,\ell_1}(\mathbf{u}_m)+\rho_{k,\ell_2}(\mathbf{u}_m)\right)\right)}\notag\\
		&=\sum_{k=1}^{K}\sum_{\ell_1=1}^{L_k}\sum_{\ell_2=1}^{L_k}\left|\left[\mathbf{A}_k\right]_{\ell_1,\ell_2}\right|\cos\left(\bar{\varrho}_{k,\ell_1,\ell_2}(\mathbf{u}_m)\right),
	\end{align}
	where $\bar{\varrho}_{k,\ell_1,\ell_2}(\mathbf{u}_m)=\arg\left(\left[\mathbf{A}_k\right]_{\ell_1,\ell_2}\right)+\frac{2\pi}{\lambda}\big(-\rho_{k,\ell_1}(\mathbf{u}_m)+\rho_{k,\ell_2}(\mathbf{u}_m)\big)$. 
	As a result, the gradient of $\Phi(\mathbf{u}_m)$ over $\mathbf{u}_m$, i.e., $\nabla\Phi(\mathbf{u}_m)=\left[\frac{\partial\Phi(\mathbf{u}_m)}{\partial x_m}, \frac{\partial\Phi(\mathbf{u}_m)}{\partial x_m}\right]^{\mathrm{T}}$, can be obtained as
	\begin{subequations}
		\label{eq40}
		\begin{align}
			&\frac{\partial\Phi(\mathbf{u}_m)}{\partial x_m}=-\frac{2\pi}{\lambda}\sum_{k=1}^{K}\sum_{\ell_1=1}^{L_k}\sum_{\ell_2=1}^{L_k}\left|\left[\mathbf{A}_k\right]_{\ell_1,\ell_2}\right|\sin\left(\bar{\varrho}_{k,\ell_1,\ell_2}(\mathbf{u}_m)\right)\notag\\
			&\times\big(-\sin\theta_{k,\ell_1}\cos\phi_{k,\ell_1}+\sin\theta_{k,\ell_2}\cos\phi_{k,\ell_2}\big),\\
			&\frac{\partial\Phi(\mathbf{u}_m)}{\partial y_m}=-\frac{2\pi}{\lambda}\sum_{k=1}^{K}\sum_{\ell_1=1}^{L_k}\sum_{\ell_2=1}^{L_k}\left|\left[\mathbf{A}_k\right]_{\ell_1,\ell_2}\right|\sin\left(\bar{\varrho}_{k,\ell_1,\ell_2}(\mathbf{u}_m)\right)\notag\\
			&\times\big(-\cos\theta_{k,\ell_1}+\cos\theta_{k,\ell_2}\big).
		\end{align}
	\end{subequations}
	Subsequently, the Hessian matrix of $\Phi(\mathbf{u}_m)$ with respect to $\mathbf{u}_m$, i.e., $\nabla^2\Phi(\mathbf{u}_m)=\begin{bmatrix}
		\frac{\partial^2\Phi(\mathbf{u}_m)}{\partial x_m^2},\frac{\partial^2\Phi(\mathbf{u}_m)}{\partial x_m\partial y_m}\\
		\frac{\partial^2\Phi(\mathbf{u}_m)}{\partial y_m\partial x_m},\frac{\partial^2\Phi(\mathbf{u}_m)}{\partial y_m^2}\end{bmatrix}$, can also be derived in a closed form, with the expressions of the matrix's elements given in \eqref{eq41}, shown at the top of this page.
	\begin{figure*}[!t]
		\begin{subequations}
			\label{eq41}
			\begin{align}
				&\frac{\partial^2\Phi(\mathbf{u}_m)}{\partial x_m^2}=-\frac{4\pi^2}{\lambda^2}\sum_{k=1}^{K}\sum_{\ell_1=1}^{L_k}\sum_{\ell_2=1}^{L_k}\left|\left[\mathbf{A}_k\right]_{\ell_1,\ell_2}\right|\big(-\sin\theta_{k,\ell_1}\cos\phi_{k,\ell_1}+\sin\theta_{k,\ell_2}\cos\phi_{k,\ell_2}\big)^2\cos\big(\bar{\varrho}_{k,\ell_1,\ell_2}(\mathbf{u}_m)\big),\\
				&\frac{\partial^2\Phi(\mathbf{u}_m)}{\partial x_m\partial y_m}=\frac{\partial^2\Phi(\mathbf{u}_m)}{\partial y_m\partial x_m}=-\frac{4\pi^2}{\lambda^2}\sum_{k=1}^{K}\sum_{\ell_1=1}^{L_k}\sum_{\ell_2=1}^{L_k}\left|\left[\mathbf{A}_k\right]_{\ell_1,\ell_2}\right|\big(-\sin\theta_{k,\ell_1}\cos\phi_{k,\ell_1}+\sin\theta_{k,\ell_2}\cos\phi_{k,\ell_2}\big)\big(-\cos\theta_{k,\ell_1}+\cos\theta_{k,\ell_2}\big)\notag\\
				&\qquad\qquad\qquad\qquad\qquad\qquad\times\cos\big(\bar{\varrho}_{k,\ell_1,\ell_2}(\mathbf{u}_m)\big),\\
				&\frac{\partial^2\Phi(\mathbf{u}_m)}{\partial y_m^2}=-\frac{4\pi^2}{\lambda^2}\sum_{k=1}^{K}\sum_{\ell_1=1}^{L_k}\sum_{\ell_2=1}^{L_k}\left|\left[\mathbf{A}_k\right]_{\ell_1,\ell_2}\right|\big(-\cos\theta_{k,\ell_1}+\cos\theta_{k,\ell_2}\big)^2\cos\big(\bar{\varrho}_{k,\ell_1,\ell_2}(\mathbf{u}_m)\big)
			\end{align}
		\end{subequations}
		\hrulefill
	\end{figure*}

	\section{Derivations of $\nabla\Gamma_{k,i}(\mathbf{u}_m)$ and $\nabla^2\Gamma_{k,i}(\mathbf{u}_m)$}
	\label{appendix_B}
	\begin{figure*}[!t]
		\setcounter{equation}{43}
		\begin{subequations}
			\label{eq44}
			\begin{align}
				\frac{\partial^2\Gamma_{k,i}(\mathbf{u}_m)}{\partial x_m^2}=&-\frac{4\pi^2}{\lambda^2}\sum_{\ell_1=1}^{L_k}\sum_{\ell_2=1}^{L_k}\left|\left[\mathbf{B}_{k,i,m}\right]_{\ell_1,\ell_2}\right|\left(-\sin\theta_{i,\ell_1}\cos\phi_{i,\ell_1}+\sin\theta_{i,\ell_2}\cos\phi_{i,\ell_2}\right)^2\cos(\bar{\omega}_{k,i,m,\ell_1,\ell_2}(\mathbf{u}_m))\notag\\
				&-\frac{4\pi^2}{\lambda^2}\sum_{\ell_3=1}^{L_k}\left|\left[\mathbf{c}_{k,i,m}\right]_{\ell_3}\right|\sin^2\theta_{i,\ell_3}\cos^2\phi_{i,\ell_3}\cos(\bar{\kappa}_{k,i,m,\ell_3}(\mathbf{u}_m)),\\
				\frac{\partial^2\Gamma_{k,i}(\mathbf{u}_m)}{\partial x_my_m}=&\frac{\partial^2\Gamma_{k,i}(\mathbf{u}_m)}{\partial y_mx_m}=-\frac{4\pi^2}{\lambda^2}\sum_{\ell_1=1}^{L_k}\sum_{\ell_2=1}^{L_k}\left|\left[\mathbf{B}_{k,i,m}\right]_{\ell_1,\ell_2}\right|\left(-\sin\theta_{i,\ell_1}\cos\phi_{i,\ell_1}+\sin\theta_{i,\ell_2}\cos\phi_{i,\ell_2}\right)\left(-\cos\theta_{i,\ell_1}+\cos\theta_{i,\ell_2}\right)\notag\\
				&\times\cos(\bar{\omega}_{k,i,m,\ell_1,\ell_2}(\mathbf{u}_m))
				-\frac{4\pi^2}{\lambda^2}\sum_{\ell_3=1}^{L_k}\left|\left[\mathbf{c}_{k,i,m}\right]_{\ell_3}\right|\sin\theta_{i,\ell_3}\cos\phi_{i,\ell_3}\cos\theta_{i,\ell_3}\cos(\bar{\kappa}_{k,i,m,\ell_3}(\mathbf{u}_m)),\\
				\frac{\partial^2\Gamma_{k,i}(\mathbf{u}_m)}{\partial y_m^2}=&-\frac{4\pi^2}{\lambda^2}\sum_{\ell_1=1}^{L_k}\sum_{\ell_2=1}^{L_k}\left|\left[\mathbf{B}_{k,i,m}\right]_{\ell_1,\ell_2}\right|\left(-\cos\theta_{i,\ell_1}+\cos\theta_{i,\ell_2}\right)^2\cos(\bar{\omega}_{k,i,m,\ell_1,\ell_2}(\mathbf{u}_m))\notag\\
				&-\frac{4\pi^2}{\lambda^2}\sum_{\ell_3=1}^{L_k}\left|\left[\mathbf{c}_{k,i,m}\right]_{\ell_3}\right|\cos^2\theta_{i,\ell_3}\sin(\bar{\kappa}_{k,i,m,\ell_3}(\mathbf{u}_m)).
			\end{align}
		\end{subequations}
		\hrulefill
	\end{figure*}
	Recall that $\mathbf{h}_k(\tilde{\mathbf{u}})=\mathbf{G}_k^{\mathrm{H}}(\tilde{\mathbf{u}})\mathbf{f}_k$. Let $w_{k,n}$ denote the $n$-th element of $\mathbf{w}_k$ and define $\zeta_{k,i,m}=\sum_{n\neq m}\mathbf{f}_i^{\mathrm{H}}\mathbf{g}_i(\mathbf{u}_n)w_{k,n}$. Armed with these, $\left|\mathbf{h}_i^{\mathrm{H}}\left(\tilde{\mathbf{u}}\right)\mathbf{w}_k\right|^2$ can be expanded as
	\begin{align}
		\label{eq42}
		\setcounter{equation}{41}
		&\Gamma_{k,i}\left(\mathbf{u}_m\right)=\left|\mathbf{h}_i^{\mathrm{H}}\left(\tilde{\mathbf{u}}\right)\mathbf{w}_k\right|^2=\left|\mathbf{f}_i^{\mathrm{H}}\mathbf{g}_i(\mathbf{u}_m)w_{k,m}+\zeta_{k,i,m}\right|^2\notag\\
		&=\sum_{\ell_1=1}^{L_k}\sum_{\ell_2=1}^{L_k}\left[\mathbf{B}_{k,i,m}\right]_{\ell_1,\ell_2}e^{\mathrm{j}\left(\frac{2\pi}{\lambda}\left(-\rho_{k,\ell_1}(\mathbf{u}_m)+\rho_{k,\ell_2}(\mathbf{u}_m)\right)\right)}\notag\\
		&+\sum_{\ell_3=1}^{L_i}\left[\mathbf{c}_{k,i,m}\right]_{\ell_3}e^{-\mathrm{j}\frac{2\pi}{\lambda}\rho_{i,\ell_3}\left(\mathbf{u}_m\right)}+\left|\zeta_{k,i,m}\right|^2\notag\\
		&=\sum_{\ell_1=1}^{L_i}\sum_{\ell_2=1}^{L_i}\left|\left[\mathbf{B}_{k,i,m}\right]_{\ell_1,\ell_2}\right|\cos\big(\bar{\omega}_{k,i,m,\ell_1,\ell_2}(\mathbf{u}_m)\big)\notag\\
		&+\sum_{\ell_3=1}^{L_i}\left|\left[\mathbf{c}_{k,i,m}\right]_{\ell_3}\right|\cos\big(\bar{\kappa}_{k,i,m,\ell_3}(\mathbf{u}_m)\big)+\left|\zeta_{k,i,m}\right|^2,
	\end{align}
	where $\mathbf{B}_{k,i,m}\triangleq\left|w_{k,m}\right|^2\mathbf{f}_i\mathbf{f}_i^{\mathrm{H}}$,  $\mathbf{c}_{k,i,m}\triangleq2w_{k,m}^*\zeta_{k,i,m}\mathbf{f}_i$, $\bar{\omega}_{k,i,m,\ell_1,\ell_2}\left(\mathbf{u}_m\right)\triangleq\mathrm{arg}\left(\left[\mathbf{B}_{k,i,m}\right]_{\ell_1,\ell_2}\right)+\frac{2\pi}{\lambda}\big(-\rho_{i,\ell_1}\left(\mathbf{u}_m\right)+\\\rho_{i,\ell_2}\left(\mathbf{u}_m\right)\big)$, and $\bar{\kappa}_{k,i,m,\ell_3}(\mathbf{u}_m)\triangleq-\mathrm{arg}\left(\left[\mathbf{c}_{k,i,m}\right]_{\ell_1,\ell_2}\right)+\frac{2\pi}{\lambda}\\\rho_{i,\ell_3}(\mathbf{u}_m)$. Therefore, the gradient vector of $\Gamma_{k,i}(\mathbf{u}_m)$ with respect to $\mathbf{u}_m$, i.e.,  $\nabla\Gamma_{k,i}(\mathbf{u}_m)=\left[\frac{\partial\Gamma_{k,i}(\mathbf{u}_m)}{\partial x_m},\frac{\partial\Gamma_{k,i}(\mathbf{u}_m)}{\partial y_m}\right]^{\mathrm{T}}$, can be derived in a closed form as
	\begin{subequations}
		\label{eq43}
		\begin{align}
			&\frac{\partial\Gamma_{k,i}(\mathbf{u}_m)}{\partial x_m}=\frac{2\pi}{\lambda}\sum_{\ell_1=1}^{L_k}\sum_{\ell_2=1}^{L_k}\left|\left[\mathbf{B}_{k,i,m}\right]_{\ell_1,\ell_2}\right|\sin(\bar{\omega}_{k,i,m,\ell_1,\ell_2}(\mathbf{u}_m))\notag\\
			&\times\left(\sin\theta_{i,\ell_1}\cos\phi_{i,\ell_1}-\sin\theta_{i,\ell_2}\cos\phi_{i,\ell_2}\right)\notag\\
			&-\frac{2\pi}{\lambda}\sum_{\ell_3=1}^{L_k}\left|\left[\mathbf{c}_{k,i,m}\right]_{\ell_3}\right|\sin\theta_{i,\ell_3}\cos\phi_{i,\ell_3}\sin(\bar{\kappa}_{k,i,m,\ell_3}(\mathbf{u}_m)),\\
			&\frac{\partial\Gamma_{k,i}(\mathbf{u}_m)}{\partial y_m}=\frac{2\pi}{\lambda}\sum_{\ell_1=1}^{L_k}\sum_{\ell_2=1}^{L_k}\left|\left[\mathbf{B}_{k,i,m}\right]_{\ell_1,\ell_2}\right|\sin(\bar{\omega}_{k,i,m,\ell_1,\ell_2}(\mathbf{u}_m))\notag\\
			&\times\left(\cos\theta_{i,\ell_1}-\cos\theta_{i,\ell_2}\right)\notag\\
			&-\frac{2\pi}{\lambda}\sum_{\ell_3=1}^{L_k}\left|\left[\mathbf{c}_{k,i,m}\right]_{\ell_3}\right|\cos\theta_{i,\ell_3}\sin(\bar{\kappa}_{k,i,m,\ell_3}(\mathbf{u}_m)).
		\end{align}
	\end{subequations}
	Similarly, the Hessian matrix of $\Gamma_{k,i}(\mathbf{u}_m)$ with respect to $\mathbf{u}_m$, i.e., $\nabla^2\Gamma_{k,i}(\mathbf{u}_m)=\begin{bmatrix}
		\frac{\partial^2\Gamma_{k,i}(\mathbf{u}_m)}{\partial x_m^2},\frac{\partial^2\Gamma_{k,i}(\mathbf{u}_m)}{\partial x_m\partial y_m}\\
		\frac{\partial^2\Gamma_{k,i}(\mathbf{u}_m)}{\partial y_m\partial x_m},\frac{\partial^2\Gamma_{k,i}(\mathbf{u}_m)}{\partial y_m^2}
	\end{bmatrix}$, can also be derived in a closed form, with the expression given in \eqref{eq44}, shown at the top of the previous page.
}

\bibliography{reference}
\bibliographystyle{IEEEtran}

\vfill
\end{document}